\def\refindent{\hangindent=1em \hangafter=1 \noindent} 

\def\ref #1;#2;#3;#4\par{\refindent #1,#2,#3,#4.\par}

\documentstyle[12pt,aasms]{article}

\begin{document}

\title{SIMULTANEOUS HST/XTE OBSERVATIONS OF 
SCO X-1}

\centerline{ }
\centerline{ }

\author{T. Kallman$^1$}

\affil{Laboratory for High Energy Astrophysics \
          NASA/Goddard Space Flight Center} 

\author{B. Boroson$^2$ and S. D. Vrtilek$^3$}

\affil{Harvard-Smithsonian Center for Astrophysics}

\begin{abstract}

Sco X-1 is the brightest extra-solar point source of X-rays, and 
may serve as a prototype for low mass X-ray binaries as 
a class.  It has been suggested that the UV and optical emission 
arise as a result of reprocessing of X-rays, and that a
likely site for such reprocessing is an accretion disk around
the X-ray source.  If UV and optical emission are enhanced 
by reprocessing of X-rays, the
X-ray variability may be manifest in UV emission:  we test this by 
using high temporal resolution UV data obtained simultaneously with 
high temporal resolution X-ray data obtained simultaneously with the 
GHRS on the Hubble Space Telescope, and with the X-ray 
Timing Explorer (XTE).  We analyze the variability 
behavior of the UV spectrum and of the X-rays, and we 
also measure the properties of the emission line profiles 
as viewed at high resolution (resolving power$\simeq$25,000) 
with the echelle gratings.
The variability behavior does not provide direct support of the 
reprocessing scenario, although the correlated variability between
UV and X-rays does not conflict with this hypothesis.  
Furthermore, the emission line profiles do not fit 
with simple models for disk emission lines.

\end{abstract}


\vskip 3.0 truein
$^1$email: tim@xstar.gsfc.nasa.gov
$^2$email: bboroson@riva.harvard.edu
$^3$email: svrtilek@cfa.harvard.edu
\vfill
\eject

\section{INTRODUCTION}

X-ray binaries remain among the most fascinating astronomical
systems.  Pronounced periodic and aperiodic variability of the
X-ray emission, on timescales ranging from milliseconds to decades,
distinguishes the X-ray binaries from all other astronomical
sources.  For example, quasi-periodic oscillations (QPOs) in the
X-rays from low mass X-ray binaries (LMXB's) (Lewin {\it et al}
1988; van der Klis, 1989) may help us understand accretion
disks about compact objects near their Eddington luminosities, disk
interactions with the magnetic fields of neutron stars, and
pulsar evolution (e.g., Lamb 1988).
Yet the current X-ray observations lack the ability to measure local
physical conditions and dynamics with high-resolution spectrometry. 
In the optical and UV wavebands we can measure variability of the
UV spectra of these sources on short timescales and at spectral 
resolution adequate to dissect accretion flows according to velocity. 
Thus, time-resolved UV spectrometry constitutes an important
discovery space for these sources.

A prime target for such observations is Sco~X-1, the
first discovered and brightest
extra-solar X-ray source and also the strongest LMXB UV source
(Willis, et al., 1978).  A prototype of the LMXBs, 
Sco~X-1 consists of a neutron star and a low mass $(\le 1 M_{\odot})$
secondary in a close ($\sim 10^{11}$ cm) 0.787d binary orbit
(Crampton {\it et al.} 1976).
Sco~X-1 is one of two ``Z-class'' sources
(displaying horizontal, normal, and flaring branches in the X-ray spectrum;
see, e.g. Lewin and van der Klis, 1994, for a review) with an identifiable 
optical counterpart.  The secondary (optical) star has a mass approximately 
1 M$_\odot$; the spectrum shows emission lines due to 
H and He II and no recognizable absorption features; if the 
mass transfer is driven by stellar evolution, then the star is 
at or near end of its main sequence age (Cowley, 1976; Cowley and 
Crampton, 1975; Gottlieb, Wright, and Liller, 1975).
The increase in QPO frequency as the source traverses the `Z' shaped 
path in the X-ray color-color diagram has been intrepreted as being due to 
an increase in mass accretion rate.  If so, the `normal branch' QPOs 
are characterized by an inverse behavior between mass accretion rate and 
X-ray luminosity.  More recently, Sco X-1 has been shown to be one of the 
sources of high frequency (kHz) QPOs (van der Klis et al., 1996; 1997).

The near-absence of stellar absorption features has slowed the interpretation of 
the optical and UV spectra of LMXBs.  The mass transfer rates necessary to 
fuel the X-ray emission (assuming 10$\%$ efficiency), together with the stellar
masses and orbital periods, suggest a similarity between LMXBs and high mass accretion 
rate cataclysmic variables (i.e. nova-like variables; Cordova and Mason, 1982).  However, the optical
and UV luminosities of LMXBs exceed those of CVs by factors 10-100.
This suggests that most of the optical and UV
luminosity ($L_{opt} \sim L_{UV} \sim 10^{-2} L_X$) comes from the
accretion flow as it is illuminated by the X-rays from the neutron
star.   

The UV observations of Sco X-1 taken prior to September 1988 by the 
International Ultraviolet Observer (IUE) satellite have been 
summarized by us in an earlier paper (Kallman, Vrtilek, and Raymond, 1990, hereafter 
Paper I), and more recent IUE data by Vrtilek, et al. (1991).
These data have been  tested for variability associated with 
orbital motion, and for correlations among the various observables.  
The continuum spectra were found to fit
to simple accretion disk atmosphere models which demonstrate that X-ray
heating dominates the outer regions of the accretion disk.  
Correlations were found among the various emission line strengths, 
and upper limits to the emission line widths provided constraints on the 
location of the line emitting material if the emission is assumed to 
come from a Keplerian disk.
IUE has shown that the UV emission line spectrum of Sco~X-1 (Paper I;
Vrtilek {\it et al.} 1991) is dominated by strong emission lines
of C IV near 1550$\AA$ and N V near 1238$\AA$ and shows
weaker lines of Si IV, O IV, O V, N IV, C III and less ionized
species.   
The UV line strengths, ratios, and profiles vary noticeably ($\sim
20\%$) on the shortest timescales ($\sim$0.5 hr) that can be probed
with IUE; but IUE does not have the time resolution to
detect variability in the UV at frequencies higher than those 
corresponding to this timescale.  If such frequencies were accessible 
observationally we could obtain information about the behavior of the
accretion flow within the Roche lobe of the primary.  

In spite of the effort devoted to understanding the X-ray variability 
behavior of LMXB's, little is known about their circumstellar environment
and evolution.  For example, estimates of the distance and 
reddening to Sco X-1 
vary by a factor of 3 (Schachter et al., 1987; Willis, et al., 1980); 
secure distance estimates are vital to the understanding of the 
energetics of the X-ray emission.  In addition, the X-rays and possible 
X-ray induced outflows from LMXB's are likely to affect the circumstellar 
environment by producing significant column densities of highly ionized 
species.  Absorption studies, using Sco X-1 as a UV light source, can test 
for these ions.  HST spectral resolution $\Delta V \sim 3
{\rm km s}^{-1}$ (HRS Echelle A), enables the dissection of the circumstellar 
environment that X-ray, optical  or IUE observations cannot do.

In this paper we describe the results of observations of Sco X-1 in the ultraviolet 
and X-ray energy bands carried out in February 1996 using the HST and 
XTE satellites.  The goals of these observations are:
(i) Search for UV variability on timescales shorter than those previously 
accessible.  The X-ray variability behavior has been extensively studied, and 
the X-ray power spectrum is known to extend to $\geq$ 10 Hz during all 
of the variability states (e.g. Hasinger and Kurster, 1990).  
According to the reprocessing hypothesis the UV 
will reflect the variability of the X-rays on some timescales.
For example, the accretion disk of Sco X-1 has a radius of about
2 lt-seconds so we expect that UV variability will be smeared on timescales 
much shorter than this; the  UV emitting region
should act as a ``low pass filter'' to the X-ray power  spectrum.  
(ii)  With HST, we can measure the interstellar column
density toward Sco X-1 using interstellar L${\alpha}$, thereby 
providing a third, independent distance estimate, which may help 
resolve the uncertainty.
The interstellar line is broader than either the
intrinsic L$\alpha$ emission from Sco~X-1  or the geocoronal line,
and is apparent in the IUE spectra. 
(iii) Sco X-1 is likely to have an effect on any nearby 
interstellar medium.  This can occur by X-ray photoionization, which will 
produce significant column densities of ions such as C IV and N V 
within a radius of $\sim$3 pc (McCray, Wright, and Hatchett, 1976).  
These will have a detectable signature in high resolution observations 
as narrow features, with widths of $\sim 0.1 \AA$.  The relative 
strengths of such features provide a measure of the ionization balance in the interstellar 
medium, which in turn can constrain the density and X-ray flux history.

In what follows we describe our attempt to test some of these ideas 
using simultaneous observations of Sco X-1 by HST and XTE.  In 
section 2 we describe the observations themselves; in sections 3 and 4
we discuss the spectral and timing analysis, respectively.  In section
5 we discuss the results and summarize the conclusions.

\vfill
\eject

\section{OBSERVATIONS AND METHODS}

\subsection{HST Data}

The data used in this analysis represents all of the available spectra of Sco
X-1 taken by the Hubble Space Telescope (HST) during 29 February, 1996 using the 
Goddard High Resolution Spectrograph (GHRS; Brandt, et al., 1990).  This was after the 
(first) HST refurbishment mission that installed COSTAR and repaired the side 1
electronics problem with the GHRS (Hartig et al., 1993).
A description of the GHRS and gratings is given by Cardelli, Ebbetts and Savage (1990).

Sco~X-1 was observed for a total of 6 satellite orbits with the GHRS.
The grating/datamode combinations we used include: (i) the G140L 
(Energy (E) resolution $\Delta E/E \sim$150 km s$^{-1}$/c) grating in 
the wavelength region 1300 -- 1580 $\AA$ (i.e. including both the Si IV $\lambda$ 1400 
and C IV $\lambda$ 1550 resonance lines) for one satellite orbit in order to test for
variability in the UV continuum and line strengths.  Data was taken in 
RAPID Mode with data readout every 0.1s. 
Owing to limitations in the capacity of the tape recorders on HST the orbit was split 
into two separate exposures of approximately 1000 sec each, separated by a 9 minute 
interval in order to read the data to the ground.
(ii) the G160M (Energy (E) resolution $\Delta E/E \sim$15 km s$^{-1}$/c) grating was used 
in the vicinity of the L $\alpha$ (1200-1233 $\AA$),
N V (1220-1253 $\AA$) and C IV (1535-1568 $\AA$) lines (in ACCUM mode) for one 
satellite orbit each to measure the damping wings of the interstellar 
L$\alpha$ absorption and hence the interstellar column,  and
to measure the profiles of the N V and C IV emission lines.
(iii)  the Echelle A grating  (in ACCUM mode) was used in the vicinity of the
N V $\lambda$1240 (order 45) and C IV $\lambda$1550 (order 36) lines for one 
satellite orbit each 
to search for narrow absorption components due to photoionized 
interstellar gas and a possible wind-blown bubble.
For all gratings the Large Science Aperture was used, which has a field of view of 
1.74\arcsec.
A journal of observations is presented in Table~1.

The data analysis procedure consisted of extraction of flux-calibrated
spectra, followed by continuum fitting and emission line
fitting.  No reddening correction has been applied to the data, although 
we do apply corrections as part of our fitting and interpretation procedure.
The data were reduced in a uniform manner using the standard                
products produced by the HST analysis pipeline.  In the case of the 
observations made in ACCUM mode these consist of the observed fluxes 
vs. wavelength, and errors vs. wavelength,  corrected for
flat field, with diode substeps added. 
We estimate a typical wavelength scale accuracy of 0.1 $\AA$ at 1300 $\AA$ for the 
G160M grating.  The standard pipeline data reduction method was used.  
GHRS exposures were not taken in FP-SPLIT mode. 
A very useful property of the GHRS is that 
the random flux errors in the data are due simply to photon
statistics and can be computed easily.  

In the case of the RAPID mode data, flux and wavelength calibration 
are not performed as part of the standard analysis.  Therefore, the absolute fluxes
we derive are uncertain by as much as 10$\%$.  
However,  the primary goal of our analysis is to search for variability
in the UV and we know of no instrumental effects which are likely to 
introduce spurious variability on short timescales.  Furthermore, 
the field containing Sco X-1 is known to be free of other nearby 
sources which could potentially contaminate the UV spectra we 
observe (Sofia et al.,1969) .  We use the large science aperture, which reduces 
the possibility of variability due to fluctuations in the satellite pointing direction.
We therefore use the raw counting rates during the RAPID mode observations
for our analysis of variability in the UV spectrum from Sco X-1.

\subsection{XTE Data}

Simultaneously with the HST observations, we observed Sco X-1 with the 
Rossi X-Ray Timing Explorer PCA (Bradt, Rothschild, and 
Swank 1993) continuously for 
3459 seconds on 1996 Feb. 28.   A maximum time resolution
of 0.016 s (16 ms) was used throughout, although for much of the analysis 
discussed in Section 3 we binned the data to 1 s resolution for 
the sake of convenience.  During these observations, an
offset pointing of 0.5 degree was used, resulting in a counting rate of 
6.3 $\times 10^5$ per 16 ms bin, or 3.94 $\times 10^7$ s$^{-1}$.  
We calculated power spectra of all 16ms data using 16 s data segments, 
and we calculated one average spectrum for each continuous data interval. 
This power spectrum clearly shows 6 Hz QPOs, characteristic of the 
horizontal branch QPOs which have previously been observed from Sco X-1
(e.g. Hasinger, 1986; Priedhorsky et al., 1986; van der Klis et al., 1985). 
Corrections for the PCA dead-time are not, as yet, sufficiently well 
understood to predict the Poisson component 
accurately. The conversion of the power in the QPO 
peaks to fractional rms amplitude depends on the derivative of the dead-time
transmission function with respect to count rate (van der Klis 1989), which 
is currently unknown. The dead time is expected to suppress the QPO amplitude more 
than the total count rate. Our raw (i.e., uncorrected for dead time)
fractional rms amplitudes are therefore lower limits to the true values. These 
could be several times as large.

\section{SPECTRAL ANALYSIS}

\subsection{Continua}

In order to make contact with previous observations of Sco X-1
(e.g. Willis, et al., 1980;  Paper I; 
Vrtilek, et al., 1991), we begin by showing the time averaged G140L spectra.
Figures 1a and 1b show the spectra in the two intervals of 
RAPID mode observations.  These spectra show many of the features 
found in IUE spectra (Willis, et al., 1980; Paper I; 
Vrtilek, et al., 1991):  strong emission lines of C IV, Si IV, and other 
lines, and a strong continuum with an approximately flat spectral
distribution.  In order further to interpret these spectra, 
we present fits to the lines and continua using assumptions 
similar to those of Paper I.

The continua were fitted by first excluding the regions containing
known strong emission lines from the reduced data (based on those 
detected in the IUE spectra in paper I), and then fitting to
a model spectrum using a least-squares procedure.  The model spectrum 
is the same as that used in Paper I.  It is
assumed to be that of an accretion disk and is given
by:

$$F_{\lambda}^{model}=(4\pi D^2)^{-1}
\int_{R_{in}}^{R_{out}}{F_{\lambda}^{local}(T(R))2\pi R dR} \eqno{(1)}$$

\noindent where $R$ is the
disk radius, $T(R)$ is the disk photospheric temperature,
$F_{\lambda}^{local}$ is the local spectrum radiated by the disk surface,
$R_{in}$ and $R_{out}$ are the disk inner and outer radii, respectively, 
and $D$ is the distance to Sco X-1.   We
ignore emission from the X-ray heated face of the companion star.  This has
been shown to be a very minor contribution to the UV light in the low mass
X-ray binary Cyg X-2 (Vrtilek et al., 1990), and Sco X-1 (Vrtilek et al., 1991; Paper I), 
although it appears to be a
significant contribution in the Her X-1/HZ Her system (Howarth and Wilson
1983).   As discussed in the following section, we are unable to detect the
orbital modulation which would be expected from a heated companion in the data
considered here.  Furthermore, we expect an illuminated companion more
nearly to resemble a single temperature black-body or stellar-type distribution,
significantly different from the disk spectral distribution; our success
in fitting to disk spectra provides {\it post hoc} justification for
neglect of companion star heating.

Theoretical calculations of the local disk spectrum are affected by the various
and uncertain heating mechanisms which may affect the disk atmosphere (see, e.g.
Shaviv, 1989) and by the problems of line blanketing and non-LTE conditions
familiar from the modelling of stellar atmospheres.  In the absence of
detailed models for the continuum spectra emitted by accretion disk
atmospheres there are two plausible choices for $F_{\lambda}^{local}$: 
normal stellar atmospheres or black-body distributions.  Studies of
cataclysmic variables have shown that neither of these 
distributions closely matches observations of CV accretion disks in
detail (Wade l984).   Many of the differences between these two cases, and
their effects on the total disk spectra have been discussed by Vrtilek, et
al. (1990), who adopt stellar atmospheres in their fitting of the spectra
of Cyg X-2.  In our models we follow Vrtilek et al. (1990) and adopt the
stellar 	atmospheres from the compilation by Kenyon (1989) when
calculating  $F_{\lambda}^{local}$.  For the purposes of modelling the UV continuum the inner
optically thin or Compton-scattering dominated regions of the disk are
unimportant (Shakura and Sunyaev, 1972; Eardley, Lightman, Payne and
Shapiro, 1978). 

The disk temperature distribution may be influenced
by a variety of factors, including the viscous release of energy by the
accreting material and heating by X-rays from the central compact object. 
Viscosity alone produces a temperature distribution $T_{acc}(R)=(3G M \dot
M/8 \pi \sigma)^{1/4} R^{-3/4}$, where $M$ and $\dot M$ are the central compact
object mass and mass accretion rate, respectively.  X-ray illumination
produces a temperature distribution $T_{x}(R)=(L_x f_x/( 4 \pi 
\sigma))^{1/4} R^{-1/2}=(\dot M c^2 \eta f_x/(4\pi \sigma))^{1/4} R^{-1/2}$, where
$L_x$ is the X-ray luminosity and $\eta$ is the accretion efficiency, which
we assume to be 0.1.  The illumination of the disk by X-rays is
parameterized by the factor 

$$f_x=(4 \pi R^2 F_{x, inc})/L_x \eqno{(2)}$$

\noindent the ratio of the X-ray flux incident
on the disk surface, $F_{x, inc}$ to the total available unattenuated X-ray
flux at that radius.  The quantity $f_x$ may be regarded as
analogous to the `Eddington factor' familiar from the study of stellar
atmospheres.  If the X-rays are unattenuated and streaming
radially outward from a point source at the center of the disk then
$f_x$=sin$\theta$, where $\theta$ is the local flaring angle of the disk. 
On the other hand, if the X-rays are nearly isotropic with the
corresponding mean intensity, $f_x\simeq 1$.  There is no {\it a priori}
reason to adopt either of these scenarios, since there is evidence that the
disks in low mass X-ray binaries are surrounded by X-ray
induced coronas (Begelman and McKee, 1982; Begelman, McKee and Shields,
1982; White and Holt, 1982).  Such coronas will have temperatures $\sim
$10$^7$ -- 10$^8$ K and optical depths to electron scattering 0.1 -- 1, and
so can isotropize the X-rays illuminating the disk.  The detailed
properties of such a corona, including its radial extent, are somewhat
uncertain, although the models of London (1984) suggest that $f_x \leq
10^{-1}$ for accretion onto a neutron star at a rate less than the
Eddington limit. 

A consequence of the different temperature dependences of the disk
temperatures produced by accretion and X-ray heating, and of the great range
of radii spanned by likely disk models, is that X-ray heating is likely to
dominate viscous heating at large radii, i.e. $R \ge R_{eq}=3GM/(2f_x\eta
c^2)=$2.2 $\times$ 10$^9$ cm
$(\eta/0.1)^{-1}(M/M_\odot)(f_x/10^{-3})^{-1}$.  Thus, even for very small
values of $f_x$, X-ray heating can determine the temperature at large radii,
and can strongly affect the UV continuum. 

The integrated spectrum given by equation (1) will be influenced by the sum
of the local spectra from the various disk radii at wavelengths less than
those which characterize the outer disk radius, $\lambda_{out}\sim hc/(k$ 
max$(T_x(R_{out}),T_{acc}(R_{out})))$.  If so, a disk which radiates locally
as a black body will have $F_{\lambda}\sim\lambda^{-7/3}$ if viscous heating
dominates and $F_{\lambda}\sim\lambda^{-1}$ if X-ray heating (with $f_x$
constant across the disk surface) dominates.  Stellar atmosphere spectra
are generally flatter than black body spectra at the corresponding
temperature.  For $\lambda\geq\lambda_{out}$, the spectrum has a
Rayleigh-Jeans slope, $F_{\lambda}\sim\lambda^{-4}$. 

In evaluating the model spectrum we assume that Sco X-1 is a 1.4 M$_\odot$
Eddington limited neutron star emitting X-rays at L=$1.9 \times 10^{38}$ erg s$^{-1}$
(the Eddington limit for 1.4 $M_\odot$),
implying a distance $D$=2 kpc, (Vrtilek, et al., 1991).  We
take the accretion efficiency $\eta$=0.1, so that $\dot M \leq$
1.9 $\times$ 10$^{18}$ gm s$^{-1}$ (corresponding to 3.05 $\times$ 10$^{-8} M_\odot
$ yr$^{-1}$.  The free parameters used to maximize
the fit to the observed spectrum include $f_x$ (assumed to be constant over
the disk surface), the mass accretion rate, and the disk outer radius,
$R_{out}$.   The observed spectrum was dereddened using the Savage and Mathis (1988) 
reddening curve and E(B-V)=0.3 (c.f. Paper I), i.e. we multiply the 
observed flux by the factor $10^{y {\rm E(B-V)}/2.5}$, where $y$ is obtained
by interpolating the Savage and Mathis curve at the desired wavelength.
The mass accretion rate is constrained to be  less than the
Eddington limit, i.e. $\dot M\leq 1$ (in units 1.9 $\times$ 10$^{18}$ gm s$^{-1}$);  
when this limit is reached  X-ray heating is the dominant energy
source for the UV emitting part of the disk.  Also, we constrain the disk outer 
radius to be less than 1 (in units of 2 light-seconds, 6 $\times$ 10$^{10}$ cm); 
this is the most probable outer radius found by Vrtilek et al. (1991). 
The values of  $f_x$  for the two G140L spectra are 3.39E-03 and 2.45E-03, 
together with the limiting  values for $\dot M$  and $R_{out}$  provide adequate fits to the
data.  The spectral slope in the UV is flatter than
$F_{\lambda}\sim\lambda^{-7/3}$ behavior expected for viscous dominated disks.  This
is consistent with the effects of X-ray heating, but is partially offset by
the fact that the UV is affected by the Rayleigh-Jeans turnover at
$\lambda_{out}\sim$7000$\AA$.   We emphasize that the continuum fitting results  provide 
strong evidence for a disk heating source in addition to viscous dissipation.
If X-ray heating were not occurring, then a viscous disk with a mass accretion rate
corresponding to the observed luminosity would fail to account for the observed 
UV flux by a factor $\simeq$16.  The average dereddened UV flux in the bandpass 
of the G140L (1260$\AA$--1550$\AA$) grating is 
$F_\lambda \simeq 3 \times 10^{-12}$ erg cm$^{-2}$ s$^{-1}$ $\AA^{-1}$, 
corresponding to a UV luminosity of 10$^{36}$ erg s$^{-1}$ if the distance to 
Sco X-1 is 2 kpc.  Although this is uncertain by a factor of $\simeq$2 owing to 
uncertainties in the reddening correction (see below), 
it still is much greater than that observed from the most luminous 
cataclysmic variable.
This may be compared with the observed flux in the 2-10 keV 
X-ray band as observed by XTE of $4.8 \times 10^{37}$ erg s$^{-1}$ (see below).

\subsection{Lines}

\subsubsection{Low Resolution Time-Averaged Spectra}

Clearly apparent in Figure 1 are the strong emission lines of C IV $\lambda\lambda$
1548, 1550, Si IV $\lambda\lambda$1394, 1403, O V $\lambda$ 1371,
and O IV $\lambda$ 1339, 1344.  In addition there are indications of other 
weaker features at 1482 and 1500 $\AA$, possibly due to O I or S I 
(e.g. Morton and Smith 1973).  We defer a discussion of the absorption
lines to a later section.

Emission line strengths were 
extracted by fitting the dereddened spectra in the vicinity
of emission lines to a linear continuum plus a gaussian line or blend of
(two) lines.  This procedure was carried out for each individual image and for
the averaged spectra for the obvious strong lines: N V
$\lambda$1240, O V $\lambda$1370, Si IV $\lambda$ 1400, and C
IV $\lambda$1550.
This procedure allows for the doublet structure of the Si IV and N V
lines. H L$\alpha$ was fitted together with the fits to N V
$\lambda$1240.  We have not included continuum absorption in our fits to 
any of the lines; a measure of the validity of this assumption can be obtained 
from the results of fitting to the high resolution data in the later 
sections. Table 2 presents the line fluxes (in units of 10$^{-11}$
erg s$^{-1}$ cm$^{-2}$), equivalent widths (in $\AA$), and the fractional errors 
(1 $\sigma$) associated with these quantities. 

The G140L grating lacks sufficient spectral resolving 
power to allow us to measure the widths of features less than $\sim$100 km s$^{-1}$, 
but we can set limits on the widths of these features.  This situation is 
similar to the results from the various IUE low resolution observations
of Sco X-1 (Paper I; Vrtilek et al., 1991), in which 
widths were less than the resolution of the IUE low resolution
short wavelength instrument.
However, we also clearly detect variability in the spectrum, both lines 
and continuum, in the $\sim$ 30 minute interval separating the G140L observations.
This is apparent from the differences in the fitting parameters for both the 
continuum and the lines in in table 2, and from Figure 1, in which the 
interval 1 spectrum (dashed curve) is overlayed onto the interval 2 spectrum 
(solid curve) in the lower panel.  This variability is comparable to the 
shortest timescale variability which had previously been sampled using IUE.

\subsubsection{Medium and High Resolution Data:  Emission Lines}

We now discuss the results of the higher spectral resolution observations
taken in ACCUM mode. Tables 3 and 4  show the results of fitting the line profiles 
obtained using the G160M and echelle gratings.  These fits were 
performed separately for each line doublet, for each grating, and for each assumed 
line shape.  The statistically significant differences between the G160M and 
echelle fits, therefore, we attribute to intrinsic variability in the lines.
The errors are the 90$\%$ confidence limits obtained by the criterion 
described by Cash (1978).
Figures 8-10 show comparisons of the fits (solid curves) with the data 
(crosses).  Clearly apparent in the observed profiles 
of C IV and N V are the doublet components at 1238.821 and 1242.804 $\AA$
and 1548.202 and 1550.774 $\AA$, respectively.  
Also apparent, particularly in the echelle spectra (Fig. 10), are the 
doublet components of the interstellar absorption lines of the 
corresponding resonance lines.  Furthermore, the absorption wavelengths
are offset from both the peak wavelengths of the emission lines, and 
from the laboratory wavelengths.  The numerical values of the offets of 
the absorption lines and the peaks of the emission lines 
from the laboratory wavelengths are given in the tables.

We have tried two different trial shapes in fitting to the emission line 
profiles.  The first is a Gaussian, which corresponds crudely to the shape 
expected from an isothermal gas, or from a turbulent medium with a velocity 
distribution which mimics a thermal distribution.  In this case, the free 
parameters are the normalizations, widths, and centroids of the various emission 
features.   We fit each observation separately, and allow the normalizations 
of the various lines within each spectrum to vary independently.  
We find in all cases acceptable fits to gaussian line profiles for the various 
lines. The centroids are offset from the laboratory wavelengths by -78 -- -177 
km s$^{-1}$ (i.e. blueshifted), and the widths are 270 -- 350 km s$^{-1}$.  
We also are able to measure the 
ratio of the doublet components (1548$\AA$/1551$\AA$ for C IV and 1237$\AA$/1242$\AA$ for N V ), 
which we find to be 1.2$^+_-$0.1 and 1.1$^+_-$0.2 
for the C IV (Fig. 8) and N V (Fig. 9) 
echelle observations, respectively.  The offsets are similar to 
those measured by Crampton et al. (1976) for the Balmer lines, which suggests comparable
contributions from orbital motion and from the systemic velocity of Sco X-1.

The second trial shape is a disk 
line, in which the broadening of the line is assumed to be dominated by 
Keplerian motion in a disk.  In this case the free parameters for each line 
are the central wavelength, the outer radius of the disk, and the distribution 
of line emissivity per unit area with radius.  This last quantity we parameterize 
according to a power law distribution, i.e. $F(R)=F_0 (R/R_0)^{-\gamma}$, where 
we choose the fiducial radius, $R_0$, to be the innermost radius of the disk, $R_{in}$,
and $F_0$ is the emissivity per unit area there.  Clearly this formulation can 
be at best a crude representation of a real disk, in which the line emissivity 
must depend on factors such as the distribution of illumination or mechanical heating 
with radius.  

The most obvious and striking prediction of the disk line model is that, 
for $\gamma \leq$2, the profile will have a minimum flux near the central wavelength,
and there will be two peaks of equal strength offset from the center of the line 
by an amount corresponding to the Keplerian speed at the outer disk edge  (if
$\gamma \geq 2$, the profile will be flat, or will have peaks at the maximum 
shift from line center; we discount this possibility in the remainder of our 
discussion).  For a disk surrounding a 1.4 M$_\odot$ object, the speed of a Keplerian 
orbit at $\sim$ 6$\times$ 10$^{10}$ cm from the object is 556 km s$^{-1}$.  Such structure 
should be easily resolvable with either the G160M or the echelle gratings, and it is 
not apparent in the data from Sco X-1.  The only possible scenario under which the disk line 
model could be considered is if the separation of the two peaks is such that one of the peaks 
coincides with the interstellar absorption, thereby rendering the appearance of the 
line to be single peaked.  Since the separation in velocity of the emission and 
absorption is $\sim$ 100 km $^{-1}$, such a scenario requires separation of the emission peaks 
which is considerably less than expected for a disk with outer radius $\sim$ 6 $\times$ 10$^{10}$ cm.
In what follows we illustrate the possibility of disk line fits by using a disk outer radius 
$R_{out}$=2.5 $\times$ 10$^{11}$ cm and an inclination sin(i)=0.32.  
This radius is comparable to the best estimates 
for the binary separation (Crampton et al., 1978), and is therefore greater than the 
likely outer radius of the disk (the ``Paczynski-Smak radius''; e.g. Frank, King and Raine, 1985).
In spite of this fact, we show in Figures 8 and 9 and in table 4, the results of our attempts to 
fit disk line profiles to the echelle spectra.  The free parameters used in these fits 
are:  $\gamma$ (the power law dependence of line emissivity with radius), 
the line centroid wavelength, and the normalization for each 
component.  We use an automated procedure to fit for these quantities, together with the values 
of $R_{out}$=2.5 $\times$ 10$^{11}$ cm and sin(i)=0.32 given above.  
As shown in figures 8 and 9, we obtain 
acceptable fits to the echelle data from these fits, although the $\chi^2$ values 
are larger than those for the gaussian fits.  

\subsubsection{Medium and High Resolution Data:  Absorption Lines}

The fitting procedures described in the previous subsection yield the strengths and widths of 
the narrow absorption components seen in the echelle spectra of figures 8 and 9.  The parameters of these 
lines are listed in tables 3 and 4.  It is clear that these lines are offset from the expected 
laboratory wavelengths by amounts ranging from -26 -- -40 km s$^{-1}$, i.e. by amounts less than 
are the emission components.  The required widths imply turbulent or thermal velocities 
30 -- 40 km s$^{-1}$.  These are greater than the $\simeq$4 -- 15 km s$^{-1}$  
expected from a thermal gas near 10$^{4}$ -- 10$^{5}$ K.  The 
equivalent widths correspond to line center optical depths of 1 -- 3 for the stronger doublet 
component, although these values differ somewhat depending on the assumed shape for the 
emission component, and are systematically lower for the G160M spectral than for the echelle 
spectra.

\subsubsection{The Lyman $\alpha$ Line}

Figure 10 shows the G160M spectrum in the vicinity of the Ly $\alpha$ line.  Clearly apparent 
are the central emission component and the broad absorption due to interstellar material.  
The central emission component is likely to be dominated by the geocoronal background.  
However, the GHRS aperture is small enough that this makes a negligible contribution 
to the flux more than $\simeq$1 $\AA$ away from line center.
We have fitted this spectrum to a Gaussian emission plus absorption by a Voigt profile.
Free parameters are: the emission component central wavelength, width, and normalization,
and the Doppler and damping parameters and optical depth of the absorption.  As was done 
for the fits to other lines, we assume a linear continuum across the spectral band of the 
instrument.  The results of the fit are as follows: $v_{0 emiss}=19.5^+_-1.5$ km s$^{-1}$, 
$\sigma_{emis}=72.05^+{0.79}_-{0.94}$  km s$^{-1}$, EW$_{emis}$=5.4$+_-0.34 \AA$,
$v_{0 abs}=74^{+9}_{-40}$ km s$^{-1}$, $\tau_{abs}=9.07^+_-0.30 \times 10^8$.  We fix the 
Doppler parameter and the damping parameter at the values expected for a 100 K gas:
$v_{thermal}=1.3$ km s$^{-1}$, $a=\Gamma/(4 \pi \Delta\nu_D)=4.65 \times 10^{-3}$.
The derived optical depth corresponds to an H I column density 
$N_{H I}=8.9^+_-0.3 \times 10^{20}$ cm$^{-2}$.  This is compatible with the value derived 
in Paper I ($\leq$ 1$\sigma$) owing to the 
large statistical errors and the much greater contamination by geocoronal Ly$\alpha$
in the IUE data.  We can estimate the contribution of geocoronal 
L$\alpha$ to our data away from line center, using the known flux and the aperture of the 
instrument; the counting rate is predicted to be 0.02 counts sec$^{-1}$ diode$^{-1}$
(Ake, 1996) near 1220 $\AA$, corresponding to a total flux  
4 $\times$ 10$^{-15}$ erg s$^{-1}$ cm$^{-2}$ $\AA^{-1}$ (Ake, 1996).
This value is comparable to the error bars already assigned to 
the flux at this wavelength, so we feel confident that the geocoronal
line makes only a minor contribution to the detected flux.  In fact, 
we have experimented with subtracting this value from the measured 
spectrum and then fitting the line, and we find that the derived line 
values differ from those cited above by much less than the quoted errors.

Combining with the relation from Diplas and Savage (1994) between 
UV reddening and HI column we would predict $E(B-V)$=0.18$^+_-$0.0061.  This is less 
than the value inferred in Paper I and by Willis et al. (1978) of $E(B-V)$=0.3,
and is similar to that suggested by Schachter, et al. (1989).  
However,  there are significant 
 uncertainties associated with measuring the H I column by fitting the 
2200$\AA$ feature using IUE data, so that we do not consider the 
quantities obtained here to be inconsistent with those from Paper I 
and from Willis et al. (1978).  Furthermore, uncertainties in the 
mean density and in the homogeneity of the interstellar gas along our 
line of sight to Sco X-1 mean that this reddening value does not 
necessarily imply a distance much less than $\sim$1.5 kpc,  
as suggested by  Knude (1989).  Recent VLBA observations have limited to the trigonometric 
parallax to Sco X-1, and set a lower limit of 1300 pc on the distance 
(Bradshaw, Fomalont, and Geldzahler, 1997).  This is consistent with
the hypothesis that Sco X-1 is Eddington limited, and that the observed 
absorption and reddening are indicative primarily on the 
density of neutral hydrogen and dust along the line of sight, rather 
than the distance.

\section{TIMING ANALYSIS}

\subsection{Variability}

Figure 2 shows lightcurves from the RAPID mode data, binned into 
10 second bins.
Panel (a) is the flux in the vicinity of the C IV line, in the 
wavelength range from 1537 to the longest wavelength accessible to the 
G140L grating, approximately  1550 $\AA$.  
Panel (b) is the flux in the 
total UV band spanned by the G140L grating, and panel (c) is the X-rays,
integrated over the entire XTE bandpass. All count rates are 
 in counts per 10 second bin.  Time is measured in seconds from 
the beginning of the X-ray observations.
The extent of the overlap between the X-rays and the two intervals of 
UV observations is clear from this figure. 
Owing to the absence of flux calibration in these data it is likely that this 
effect is due to gain drifts associated with changes in conditions during the satellite 
orbit.  Therefore, in what follows we have removed this trend in the data 
by subtracting a linear fit to the mean counting rate in both the UV 
wavelength bands shown here.  As we will show, the residual flux shows variability 
on timescales much shorter than those associated with the satellite orbit, and 
which therefore are not likely to be associated with gain changes in the 
GHRS.

The mean counting rates and fluctuations associated with the data 
in figure 2 are as follows:  
In the C IV line the mean counting rate is 124.6 s$^{-1}$ 
for the first observation interval,
104.5 s$^{-1}$ for the second interval.  In the continuum the 
rates corresponding quantities are 1773 s$^{-1}$ and 1403 s$^{-1}$, 
respectively, while in the X-ray band the counting rate is 
6.384 $\times$ 10$^5$ s$^{-1}$ during the entire interval shown in figure 2.  
As a comparison, the fractional rms variation we derive from
the various observed datasets 
(using the definition given by Lewin, Van Paradijs and Van der Klis, 
1989) is 0.45, 0.49, 0.089, 0.090, and 0.0021, for the two line 
intervals, the two continuum intervals, and the X-rays, respectively,
Since the errors are dominated by counting statistics for all of these 
data, the signal to noise ratio is $\simeq$1 -- 1.5 for all the 
observations on the shortest timescales, assuming that most of the 
variance occurs on these timescales.

More detail about the variability is shown in the power spectra, in Figure 3.  
Panel (a) shows the spectrum in the C IV line, panel (b) in the UV
continuum, and panel (c) in the X-rays.  Panels (a) and (b)
here represent the average between the two G140L observation intervals; 
all have been binned onto evenly spaced logarithmic frequency intervals, 
and the error bars shown represent the dispersion in the values within that 
interval. The normalization has been chosen according to the prescription of Leahy
et al., (1983), with the consequence that noise due to counting statistics 
has a level of 2.  Panel (a) shows a clear excess of power, by a factor $\sim$3, 
at frequencies below $\sim$ 0.1 Hz (i.e. timescales $\geq$10 seconds), 
relative to the noise level.  In the UV 
continuum and X-rays this same behavior is apparent, and the excesses are factors of 
$\sim$100, and 5 $\times$ 10$^3$, respectively.  This shows that there is a 
clear excess in variability in all the datasets at low frequencies, relative 
to the noise level predicted by counting statistics, in contrast to 
the estimates in the previous paragraph.   We have also examined the UV power spectra 
from the two intervals of data separately.  The power spectra 
show a $\simeq 20\%$ difference in the power level of the low frequency noise between 
the two intervals, with no detectable difference in the shape or high frequency behavior
between the two.

The X-ray power spectrum shows a clear 
detection of power at all frequencies, above that expected from noise alone.
Although we restrict ourselves here to frequencies accessible to both the HST 
and XTE instruments, the XTE data extends to frequencies $\simeq$60 Hz with adequate statistics 
to detect power at all frequencies.  Such spectra show clear $\sim$6 Hz QPOs, 
demonstrating that Sco X-1 was on the normal branch during our observations.  
Fits to the observed spectrum indicate a flux in the 2-10 keV band of 
$1.08 \times 10^{-7}$ erg cm$^{-1}$ s$^{-1}$ averaged over the entire interval 
shown in figure 2.

Figure 4 shows the autocorrelation functions (ACF) for the various light curves:  
the C IV line (panel a), the entire UV band (panel b), and the X-ray band (panel c). 
It is clear that the UV continuum and X-ray datasets
have correlated variability around zero lag.  Here and in what follows the  various 
correlation functions are calculated using the definitions and the subroutines 
of Press et al. (1986), i.e., the the discrete correlation function of two time series $g(t)$ and
$h(t)$ is

$${\rm Corr}(g,h)_j \Longleftrightarrow G_k H_k^*$$

\noindent where $g_j \Longleftrightarrow G_k$, $h_j \Longleftrightarrow H_k$
$H_k^*$ is the complex conjugate of $H_k$ and $\Longleftrightarrow$ 
denotes a discretely sampled Fourier transform pair.  The correlation functions 
are also normalized by dividing by the geometric mean of sums of the variance of 
each member;  this is the normalization adopted by Leahy, et al., (1983).

The fractional amplitudes of the 
zero lag modulation are 28$\%$ and 7$\%$, respectively, for the two datasets.
The full width at half maximum of the X-ray ACF is $\simeq$ 150 sec, and is indicative 
of the excess 
power in the X-rays at frequencies less than 0.1 Hz.  The full width at half 
maximum for the UV continuum is considerably greater, $\sim$300 s, and is likely 
to be affected by the window function introduced by the $\simeq$1000s observation intervals.  

Figure 5 shows the cross correlation functions (CCFs) of the various datasets:  UV line vs. X-ray, 
UV continuum vs. X-ray, and UV line vs. UV continuum.  In this case the 
cross correlations between the X-rays and the two intervals of UV data were calculated 
separately and then averaged.  Apparent in these figures are significant negative 
correlations at positive lags (UV relative to X-rays) of approximately 100 sec in 
the UV vs. X-ray panels.  The amplitude of this anticorrelation is -0.05.
The apparent sharp rise in cross correlation at larger positive lags, and the residual
negative correlation at smaller lags, are consistent with the effects of the 
window function and counting statistics (as demonstrated in the next section).

We consider this
convincing evidence for (anti) correlated variability, and for a detectable lag between the 
X-ray variability and the UV continuum.  The cross correlation of X-rays with the UV line 
does not have sufficient statistical significance to allow similar inferences to be drawn
(the minimum between 50 and 100 second lag is approximately 1$\sigma$)
although there is no obvious indication that the UV line and continuum differ in their 
variability.  Complementary evidence of this comes from the cross correlation of 
the UV line and continuum, which resembles the autocorrelation functions 
of the two UV datasets.  This is consistent with no detectable variability of the ratio of 
lines to continuum.

Fundamental to the reprocessing model for UV emission is the assumption that the 
UV and X-rays are connected by an integral equation.  If this equation is assumed to 
be linear, then the kernel, also known as the response function (Blandford and McKee, 1979), 
can in principle be obtained by inverting the equation in Fourier space.  
This procedure has been discussed at length in the context of active galactic 
nuclei (e.g. Edelson and Krolik, 1988; Krolik and Done, 1995; Done and Krolik, 1996; 
and references therein), 
and is subject to some uncertainty owing to the fact that such inversion procedures 
inevitably introduce noise into the result.  
We have attempted to apply this procedure to our data using the 
technique of regularized linear inversion (Krolik and Done, 1995; Done and Krolik, 1996), 
using a computer code kindly supplied by J. Krolik.  This procedure requires the
introduction of a default model for the response function, which we have taken 
to be constant as a function of lag.  Another parameter which affects the 
results of this procedure is the ratio of the importance of $\chi^2$ relative to
the regularization condition; we find that a value of 0.05 for this parameter gives
an acceptable value $\chi^2$=24.7.
An advantage of this procedure over 
others which have been used to solve similar problems is that it provides a 
relatively straightforward estimate of errors.  Figure 7 shows the response function 
obtained by this procedure, together with the estimated errors as a function of 
positive lag.  This clearly shows the same features of both the observed and 
simulated cross-correlation functions:  a negative response between UV and X-rays
near 100 seconds lag, together with an increase at larger lags.  These features 
are statistically significant (the depth of the minimum differs from the value at 
zero lag by 2.5 $\sigma$) and provide further confirmation of the 
results of the previous sections.

\subsection{Simulations}

In order better to understand the variability in both UV and X-ray, and the correlated 
behavior, we have simulated the data using shot noise models.  For the X-rays, these 
are chosen for the purpose of convenience; we do not regard them as necessarily being a 
plausible physical model for the variability.  Rather, they adequately reproduce the gross 
features of the power spectrum and the autocorrelation.  Therefore they 
serve as a convenient driving function for use in probing the correlated UV variability.
The shot noise model we adopt has exponential shots, with characteristic 
decay time 100 seconds.  The shots are generated randomly in time according to a Poisson 
distribution with characteristic interval 1 s$^{-1}$.  In addition, there is a 
component which we model as being constant in time.  The relative contribution of 
shot component and the non-shot (constant) component to time series is 30:1.  
The time series produced by these two components is also modified by the addition 
of noise associated with counting statistics.  Since the signal to noise ratio 
of the X-ray time series is very large, counting statistics have a negligible effect 
on the power series, ACF, and CCF.  However, in order to account approximately for the  
X-ray variability at the highest frequencies of interest here, we have 
artificially amplified the noise by a factor of 5 relative to what would be produced 
by counting statistics alone.  

Using this simulated X-ray time series we obtain a power spectrum which resembles the 
observed spectrum in its overall shape, and agrees in the magnitude of the power to 
within a factor of 2.
The long shots have the effect of producing the relatively 
large power observed in X-rays at frequencies less than 10$^{-2}$ Hz, while the amplified noise
produces the power observed at higher frequencies.  
The simulation succeeds in reproducing the exponential falloff of the autocorrelation function 
for lags less than about 200 s.  This is apparent from the third panel of Figure 6.

The key goal in constructing these simulations is modelling the UV power spectra and 
correlation functions.  In doing so, we have tried various possible assumptions about the 
UV response to X-rays.  These include:  uncorrelated UV variability, correlated UV variability
with no lag, correlated UV variability with a lag (possibly negative), and anti-correlated 
UV variability with or without lags.   Discussion of these is simplified if we refer 
to them according to simple mnemonics:  C, U, and A refer to correlated, uncorrelated, and 
anticorrelated, respectively, and L and N refer to lagged and unlagged.  

Although we will not display here all the results 
of such experiments, we will briefly describe some of the conclusions, many of which 
are obvious in retrospect.   First, simulated UV  power spectra assuming a pure secular 
decrease in the flux, plus counting statistics, can reproduce the observed power spectra
shown in Figure 3. This corresponds to model UN in the notation of the previous
paragraph.   However, such simulations do not adequately reproduce either the 
autocorrelation or UV-X-ray cross corrrelations.  Instead, we model the UV 
timeseries as the sum of two components: one which is constant in time, and one which 
responds to the shot noise component of the X-rays 
(with an amplitude which may be negative), and which may be lagged 
relative to the X-rays.  The relative contributions of these two components to the UV lightcurve
is set by the amplitude, which we denote $A$.  The count rate obtained in this way 
is then corrected for the effects of noise due to counting statistics, again assuming 
that the noise follows a Gaussian distribution.  For $A$=0, the UV lightcurve will therefore 
be pure noise at the observed mean counting rate (model UN).  For $A <$0 there will be additional 
variable component which will be anticorrelated with the shot noise part of the 
X-rays (model AN), and for $A >$0 the UV and X-rays will be correlated model (model CN).  

Following the results of the 
previous section, we take $A$=-0.04, and assume a lag of 100 s.  This corresponds to 
model AL in the above notation.  The cross correlation 
produced by this simulation is shown in Figure 6.  Although the detailed shapes differ from 
the observed data, this figure does reproduce the gross features of the observed CCF, namely
the negative correlation near 100 s, relatively constant values at smaller lags, and 
the steep rise at longer lags.  Experiments verify  that the strength and location of the 100s
feature are directly proportional to the values of $A$ and lag time used in the simulation.
Also, the behavior at larger positive lags is similar to that observed, and experiments show that
this depends solely on the window function for the observations.   The width of the 100s feature
is narrower than observed (40$^+_-$10 s. in the simulation, FWHM, vs. 110$^+_-$30 s. in the data), 
suggesting that there is additional 
short timescale variability in the UV which we have not adequately modelled.   However, 
the qualitative agreement between the observed and simulated CCFs confirms the existence of 
lagged, anticorrelated variability between the X-ray and UV continuum.
The other models fail to account for the minimum in the UV continuum/X-ray anticorrelation
at 100 second lag; model CL produces a maximum rather than a minimum, and 
models UL and  UN produce no features in the CCF.

\section{Discussion}

The cross correlation between UV and X-rays provides evidence for inverse behavior 
between the X-rays and the UV lines and continua.  This contrasts with the simplest
reprocessing models, in which the lines and continua arise from an X-ray heated 
accretion disk atmosphere.  Such a scenario would predict a positive correlation 
between X-rays and UV, with a lag associated with the combined effects of 
light travel time across the disk and the local timescale for 
reprocessing to occur on the disk surface.  Plausible values for these timescales 
are $\sim$1 s and $\sim$10 s (London and Cominsky, 1984), respectively.  
The signal to noise ratio in our data is too large to detect correlated 
variability on these timescales, so we cannot rule out the validity of this 
scenario.  Furthermore, the energetic arguments in favor of reprocessing remain 
valid, and so we favor models for Sco X-1's UV production in which reprocessing 
remains the primary mechanism.

The inverse behavior we observe on longer timescales suggests that increases 
in the X-ray luminosity impede production of UV photons, with a lag 
of approximately 100 seconds.  
This could occur either when X-rays  ionize the gas in the accretion 
disk atmosphere to a level where it emits UV with low efficiency, or by affecting 
the disk structure so that a small fraction 
of the X-rays are intercepted by the disk.  
In either case a requirement is that the timescale be comparable to the 
observed lag.  The timescale which affects reprocessing on a microscopic scale 
is the photoionization timescale, which can be written
$t_{phot}\simeq 2 \times 10^{-3} \varepsilon_{10}^{-1} R_{11}^2 L_{38}^{-1} \sigma_{20}^{-1}$ s,
where $\varepsilon_{10}$ is the mean ionizing photon energy in units of 10 keV, 
$R_{11}$ is the distance from the X-ray source in units of $10^{11}$ cm,
$L_{38}$ is the X-ray luminosity in units of $10^{38}$ erg s$^{-1}$, 
and $\sigma_{20}$ is the photoionization cross section in units of $10^{-20}$ cm$^{-2}$.
This is likely to be short compared to the observed lag, and the corresponding heating 
time is shorter still.  The timecale for changes in disk density structure to propogate 
inward is $t_{viscous}\sim t_{freefall}/\alpha$, where $\alpha$ is the disk viscosity parameter, 
with a value that is in the range $10^{-2}$ -- 1, and the freefall timescale is
$t_{freefall}\simeq 1.9 \times 10^3 R_{11}^{3/2}$ s.  A lag of the observed magnitude would 
occur if the X-rays affected the disk structure in such a way as to impede the reprocessing,
for example by decreasing the disk scale height.  If this occured a radius greater than 
that where most of the reprocessing occurs, then the lag would arise as a result of the 
inward propogation of the affected region; the propogation distance would be 
$\sim 10^{10}/\alpha$cm, comparable to the disk radius.  The change in disk structure could 
be the result of X-ray ionization which ionizes or evaporates the outer disk layers.
An alternative is that the X-rays affect the mass supply into the disk, which would occur 
if the mass transfer takes place as the result of an X-ray excited wind.  
However, the naive expectation for such a scenario is that it produces a positive 
feedback, which would produce positively correlated UV/X-ray variability, 
in contrast to what we observe.

Constraints on the density in the line forming regions 
come from the relative strengths of the various emission 
lines.  From the doublet ratios we infer that the emitting gas must be at least marginally 
optically thick in the resonance lines of C IV and N V.  
The  O IV $\lambda$1340, and O V $\lambda$1370 lines are subordinate lines, and 
so require either that the emitting ions be in an excited state, or that excitation 
occur from the ground state through a dipole-forbidden transition.  The 
former case requires a combination of high gas density and optical depth in the 
resonance line leading to the level from which excitation can occur.  This can 
be expressed as  $n\tau \geq 10^{16}$ cm$^{-3}$.   Such densities and optical 
depths are predicted to occur in X-ray heated disk atmospheres 
(Ko and Kallman, 1995). 

The line profiles provide direct constraints on the dynamics in the emitting 
region.  From Gaussian fits to the echelle spectra (Fig. 10)  we infer that the gas 
velocity dispersion is less than the Keplerian speed expected at the outer disk
edge.  Fits to models for emission lines broadened by Kepler motion
in an accretion disk require a somewhat contrived spacing for the two
emission peaks expected from such a model, implying a lower Kepler 
velocity at the outer disk edge than would be predicted based on the 
other properties of the system.  One scenario which might be consistent 
with the line profiles is emission in a turbulent region associated 
with the interaction of the accretion stream with the outer disk edge.
If so, the line centroid energy would be expected to vary with orbital 
phase by an amount which reflects the orbital motion of the emission 
region about the center of mass of the system, and which could be 
as large as $\sim$100 km s$^{-1}$.   Further observations with resolution 
comparable to those used here are needed to test this possibility.

If the blueshifts of the emission lines are due in part to 
orbital motion, then it is possible that the absorption line gas is associated 
with Sco X-1.  However, the radial velocity curve of Crampton et al. (1976) 
suggest the systemic velocity is $\simeq$150 km s$^{-1}$, which is 
still greater than the absorption line velocities of 25-40 km s$^{-1}$.
Since narrow UV absorption features are unlikely to be associated 
with gas falling into Sco X-1, we consider it most probable that the 
gas responsible for the absorption features is not associated 
with Sco X-1, either through its origins or its present dynamics.
On the other had, on the basis of the data presented here we cannot
rule out this possibility with certainty.
The column densities implied by the C IV line strength is
$N\simeq 2.6 \times 10^{16} \tau T_4^{1/2} x^{-1} {\rm cm}^{-2}$, 
where $\tau$ is the line center optical depth, $T_4$ is the gas temperature
in units of $10^4$ K, and $x$ is the ion fraction. This is comparable 
to that expected from a bubble swept up by a wind with a speed 
10$^7$ cm s$^{-1}$ and a mass loss rate 10$^{18}$ gm s$^{-1}$ 
expanding into a medium with an ambient density of 1 cm$^{-3}$
(Castor, McCray and Weaver 1975).

Acknowledgements:  We thank M. Van der Klis, J. Swank, K. Mukai, J. Krolik 
and the referee for advice and 
useful suggestions.  Support for this work was provided in 
part by a  NASA grant through the Hubble Space Telescope Guest Investigator
Program.

\vfill
\eject

\centerline{REFERENCES}

\ref Ake, T., 1996;;;private communication\par
\ref Begelman, M. C., McKee, C. F., and  Shields, G.A., 1983; Ap. J.; 271;70\par
\ref Begelman, M. C., and McKee, C.F., 1983; Ap. J.; 271; 89\par  
\ref Blandford, R. and McKee, C., 1982; Ap. J.; 255; 419\par
\ref Bradshaw, C. F., Fomalont, E. B., and Geldzahler, B. J., 1997; Ap. J.; 484; L55\par
\ref Canizares, C., et al., 1975; Ap. J.; 197; 457\par
\ref Cordova, F.A., and Mason, K.O., 1982a, in;; Accretion Driven Stellar X-ray 
  Sources; ed. W.H.G. Lewin and E. P. J. van den Heuvel , Cambridge University Press, Cambridge\par
\ref Cowley, A., and Crampton, D., 1975; Ap. J. Lett.; 201; L65\par
\ref Cowley, A., 1976, in; Eighth Texas Symposium on Relativistic Astrophysics;;
New York Academy of Sciences, M. D. Papagiannis, ed\par
\ref Crampton, D., Cowley, A., Hutchings, J.B., and Kaat, C., 1976; Astrophys.
J.; 207; 907\par
\ref Diplas, A., and Savage, B.; 1994; Ap. J.; 427; 274\par
\ref Done, C., and Krolik, J., 1996; Ap. J; 463; 144\par
\ref Eardley, D., Lightman, A., Payne, D., and Shapiro, S., 1978; Ap. J.; 224;
53\par 
\ref Edelson, R. and Krolik, J., 1988; Ap. J.; 333; 646\par
\ref Frank, J., King, A., and Raine, D., 1985; `Accretion Power in Astrophysics';;Cambridge
University Press\par
\ref Giacconi, R. et al. 1962; Phys. Rev. Lett.; 9; 439\par
\ref Gottlieb, E.W., Wright, E.L., and Liller, W., 1975; Ap. J. Lett.; 195;
L33\par
\ref Hartig, G., Crocker, J, Ford, H., 1993; SPIE; 1945; 17\par
\ref Hasinger, G., Priedhorsky, W. C., and Middleditch, J., 1989; Ap. J.; 337;
843\par 
\ref Howarth, I.D., and Wilson, R.E., 1983; M.N.R.A.S.; 202; 347\par
\ref Kallman, T.R., and McCray, R. A., 1982; Ap. J. Supp.; 50; 263\par
\ref Kallman, T.R., Vrtilek, S., and Raymond, J., 1991; Ap. J.; 370; 717\par
\ref Kenyon, S., 1989;;;private communication \par
\ref Krolik, J. and Done, C., 1995; Ap. J.; 440; 166\par
\ref Ko, Y., and Kallman, T.R., 1994; Ap. J.; 431; 273\par
\ref Knude, J., 1987; Astr. Ap.; 171; 289\par
\ref Leahy, D., et al., 1983; Ap. J; 266; 160\par
\ref Lewin, W., Van Paradijs, J.., and Van der Klis, M., 1989; 
Space Science Reviews; 46; 273\par
\ref London, R., 1982; in Cataclysmic Variables and Low Mass
X-ray Binaries; J. Patterson and D. Lamb, eds.;\par
\ref London, R., McCray, R., and Auer, L., 1981; Ap. J.; 243; 970\par
\ref Milgrom, M., 1976; Ap. J.; 207; 902\par
\ref Morton, D., and Smith, W., 1973; Ap. J. Supp.; 26; 333\par
\ref Press, W., Flannery, W., Teukol\-sky, S., and Vet\-ter\-ling, W., 1986;
Numerical Recipes;; Cambridge University Press\par
\ref Priedhorsky, W., et al., 1986; Ap. J.; 306; L91\par
\ref Raymond  J.C., and Smith, B. H., 1977; Ap. J. Supp.; 35; 419\par 
\ref Ruderman, M., Shaham, J., and Tavani, M., 1989; Ap. J.; 336; 507\par 
\ref Savage, B.D., and Mathis, J.S., 1979; Ann. Rev. Ast. and Ap.; 17; 73\par
\ref Schachter, J., Filippenko, A., V., and Kahn, S.M., 1989; Ap. J.; 340; 1049\par
\ref Shakura, N.I., and Sunyaev, R.A., 1973; Astr. and Ap.; 24; 337\par  
\ref Sofia, S., Eichhorn, H., and Gatewood, G., 1969; A. J.; 74; 1\par
\ref Vrtilek,S.D., Penninx,  W.,  Raymond, J.C., Verbunt, F., Hertz, P., 
Wood, K., Lewin, W.H.G. and Mitsuda, K.  1991; ApJ; 376; 278. \par  
\ref van der Klis et al., 1985; IAU Circ; 486;;\par
\ref van der Klis et al., 1996; Ap. J.; 469;1;\par
\ref van der Klis et al., 1997; Ap. J.; 481;97;\par
\ref Wade, R. A., 1984; M.N.R.A.S.; 208; 381\par
\ref White, N.E., and Holt,
S.S., 1982; Ap. J.; 257; 318\par
\ref White, N.E. {\it et al.}, 1986;
M.N.R.A.S.; 218; 129\par
\ref White, N.E., Peacock, A., and Taylor, B.G. 1985;
Ap. J.; 296; 475\par
\ref White, N.E., Stella, L., and Parmar, A., 1988; Ap.
J.; 324; 363\par
\ref Willis, A., et al., 1980; Ap. J.; 237; 596\par 

\vfill
\eject

\topskip=1truecm

\raggedbottom

\abovedisplayskip=3mm

\belowdisplayskip=3mm

\abovedisplayshortskip=0mm

\belowdisplayshortskip=2mm

\normalbaselineskip=12pt

\normalbaselines

\centerline{Table 1}
\vskip 1.0cm

\centerline{Journal of Observations}

$$\vbox{\tt \halign {\hfil #\hfil && \quad \hfil #\hfil \cr
interval&Instrument&Start (UT)&Stop (UT)&$\lambda_{min}(\AA)$&$\lambda_{max}(\AA)$\cr
1&GHRS G140L RAPID& 50142.80780448&50142.81888521&1260&1550\cr
2&GHRS G140L RAPID& 50142.82586003&50142.83725326&1260&1550\cr
3&GHRS G160M ACCUM&50142.74333833&50142.77154290&1210&1260\cr
4&GHRS G160M ACCUM&50142.68090778&50142.74224169&1530&1570\cr
5&GHRS G160M ACCUM&50142.61785222&50142.67916298&1205&1235\cr
6&GHRS echelle A ACCUM&50142.87574574&50142.90646767&1236&1244\cr
7&GHRS echelle A ACCUM&50142.94247028&50142.97319221&1545&1555\cr
8&XTE&50142.800781250&50142.840824074&12.4&0.62\cr
}}$$

\vfill

\eject

\topskip=1truecm

\raggedbottom

\abovedisplayskip=3mm

\belowdisplayskip=3mm

\abovedisplayshortskip=0mm

\belowdisplayshortskip=2mm

\normalbaselineskip=12pt

\normalbaselines

\centerline{Table 2}
\vskip 1.0cm

\centerline{Gaussian line fits}

$$\vbox{\tt \halign {\hfil #\hfil && \quad \hfil #\hfil \cr
obs&$\lambda$&$F_{\rm line}$&$F_{\rm continuum}$&W&EW&\cr
&(\AA)&(erg cm$^{-2}$ s$^{-1}$)&(erg cm$^{-2}$ s$^{-1}$ 
\AA$^{-1}$)&(\AA)&(\AA)&\cr
G140L 
1&1370&1.08E-12$\pm$4.03E-14&2.48E-13$\pm$1.38E-14&20.7&4.3$\pm$0.3&\cr
G140L 
1&1395&2.73E-12$\pm$2.98E-14&2.24E-13$\pm$1.21E-14&31.5&12.2$\pm$0.7&\cr
G140L 
1&1549&2.73E-12$\pm$2.98E-14&2.24E-13$\pm$1.23E-14&31.5&12.2$\pm$0.7&\cr
G140L 
1&1341&3.87E-12$\pm$4.67E-14&1.81E-13$\pm$3.77E-14&7.4&21.4$\pm$4.5&\cr
G140L 
1&1480&3.87E-12$\pm$4.67E-14&1.81E-13$\pm$3.42E-14&7.4&21.4$\pm$4.1&\cr
G140L 
1&1500&4.09E-13$\pm$6.22E-14&2.71E-13$\pm$6.80E-15&9.8&1.5$\pm$0.2&\cr
G140L 
2&1370&8.86E-13$\pm$3.14E-14&1.93E-13$\pm$1.08E-14&20.7&4.6$\pm$0.3&\cr
G140L 
2&1395&2.36E-12$\pm$2.32E-14&1.75E-13$\pm$9.93E-15&31.5&13.5$\pm$0.8&\cr
G140L 
2&1549&2.36E-12$\pm$2.32E-14&1.75E-13$\pm$9.98E-15&31.5&13.5$\pm$0.8&\cr
G140L 
2&1341&3.39E-12$\pm$3.64E-14&1.41E-13$\pm$3.73E-14&7.4&24.1$\pm$6.4&\cr
G140L 
2&1480&3.39E-12$\pm$3.64E-14&1.41E-13$\pm$2.90E-14&7.4&24.1$\pm$5.0&\cr
G140L 
2&1500&3.27E-13$\pm$4.95E-14&2.10E-13$\pm$5.52E-15&9.2&1.6$\pm$0.2&\cr
}}$$

\vskip 2.0cm

\vfill

\eject

\topskip=1truecm

\raggedbottom

\abovedisplayskip=3mm

\belowdisplayskip=3mm

\abovedisplayshortskip=0mm

\belowdisplayshortskip=2mm

\normalbaselineskip=12pt

\normalbaselines

\centerline{Table 3}
\vskip 1.0cm

\centerline{Gaussian line fits}

$$\vbox{\tt \halign {\hfil #\hfil && \quad \hfil #\hfil \cr
line&instrument&$v_{\rm emiss}$&$W_{\rm emis}$&$EW_{\rm emis}$&$v_{abs}$&$W_{\rm abs}$&$EW_{\rm abs}$&$\chi^2/\nu$\cr 
&&(km/s)&(km/s)&(\AA)&(km/s)&(km/s)&(\AA)\cr
C\,IV&G160M&-77.9$_{-1.9}^{+11.6}$&349$_{-1.4}^{+0.3}$&23.5$_{-0.7}^{+2.3}$&-27.5$_{-19.4}^{+0}$&38.6$_{-0.5}^{+23.8}$&1$_{-0.3}^{+0.6}$&2263/2000\cr
&ech&-132$_{-7.8}^{+9.7}$&274$_{-7.4}^{+7.1}$&24.2$_{-3.1}^{+3.5}$&-37.2$_{-3.9}^{+5.8}$&46.5$_{-2.2}^{+0.1}$&1.6$_{-0.2}^{+0.2}$&1612/2000\cr
N\,V&G160M&-177$_{-4.8}^{+7.3}$&354$_{-5.3}^{+5.3}$&23.1$_{-1.3}^{+1.5}$&-34.1$_{-12.1}^{+55.7}$&45$_{-30.1}^{+28}$&0.2$_{-0.2}^{+0.2}$&7984/2000\cr
&ech&-117$_{-60.6}^{+0}$&246$_{-0.7}^{+20.7}$&13.1$_{-4.2}^{+2.1}$&-26.9$_{-19.4}^{+2.4}$&31.4$_{-0.7}^{+25.1}$&0.5$_{-0}^{+0.9}$&2595/2000\cr
}}$$

\vskip 2.0cm

\centerline{Table 4}
\vskip 1.0cm

\centerline{{\em disk line} model fits}

$$\vbox{\tt \halign {\hfil #\hfil && \quad \hfil #\hfil \cr
line&instrument&$v_{\rm emiss}$&$W_{\rm emis}$&$EW_{\rm emis}$&$v_{\rm abs}$&$W_{\rm abs}$&$EW_{\rm abs}$&$\chi^2/\nu$\cr
&&(km/s)&(km/s)&$(\AA)$&(km/s)&(km/s)&$(\AA)$&\cr
C4&ech&-116.7&387.7$_{-120.8}^{+106.3}$&66.5$_{-19.8}^{+8.8}$&-39.1$_{-1.9}^{+3.9}$&43.3$_{-3}^{+3.2}$&2$_{-0.3}^{+0.3}$&1915/2000&\cr
n5&ech&-102&484.5$_{-239.2}^{+217.8}$&29.7$_{-16.4}^{+11.1}$&-31.7$_{-300.6}^{+300}$&24.2$_{-0.3}^{+13}$&0.7$_{-0.1}^{+0.4}$&3207/2000&\cr
}}$$

\vfill

\eject

\centerline{Figure Captions}

\noindent{\underbar {Figure 1}} Time averaged spectra taken by the G140L grating during the two
intervals of observation, corresponding to those described in table 1.  
Upper panel shows first interval, second panel shows second 
interval. Fluxes are in units of erg cm$^{-2}$ s$^{-1}$ $\AA^{-1}$.
The bottom panel show the logarithm of the ratio of the two intervals.

\noindent{\underbar {Figure 2}} The series from the G140L RAPID mode observations and from XTE.
Vertical axis is counts per 0.1 s time bin.  Horizontal axis is time relative to the start of the 
X-ray observation. 

\noindent{\underbar {Figure 3}}  Power spectra in the 3 energy bands corresponding to figure 2.

\noindent{\underbar {Figure 4}} Autocorrelation functions for the time series in the 3 energy 
bands corresponding to figure 2.  

\noindent{\underbar {Figure 5}} Cross correlation functions between the time series in the 3 energy 
bands corresponding to figure 2.

\noindent{\underbar {Figure 6}} Simulated cross correlation functions between the time series in the 3 energy 
bands corresponding to figure 2.

\noindent{\underbar {Figure 7}} Response function calculated using the regularized linear inversion 
algorithm.  Units of response are s$^{-1}$.

\noindent{\underbar {Figure 8}} Echelle spectrum in the vicinity of the C IV line (crosses).  Data has been binned into 
1 $\AA$ bins, and the error bars reflect the dispersion about the mean value of the data within that bin.
Solid curve is best fit disk line spectrum.   Fluxes are in units of erg cm$^{-2}$ s$^{-1}$ $\AA^{-1}$.

\noindent{\underbar {Figure 9}} Echelle spectrum in the vicinity of the N V line (crosses).  Data has been binned into 
1 $\AA$ bins, and the error bars reflect the dispersion about the mean value of the data within that bin.
Solid curve is best fit disk line spectrum.   Fluxes are in units of erg cm$^{-2}$ s$^{-1}$ $\AA^{-1}$.

\noindent{\underbar {Figure 10}} G160M spectrum in the vicinity of the L$\alpha$ line (crosses).  
Solid curve is best fit damped absorption line spectrum plus a narrow gaussian emission.
 Fluxes are in units of erg cm$^{-2}$ s$^{-1}$ $\AA^{-1}$.

\input epsf           
\epsfbox{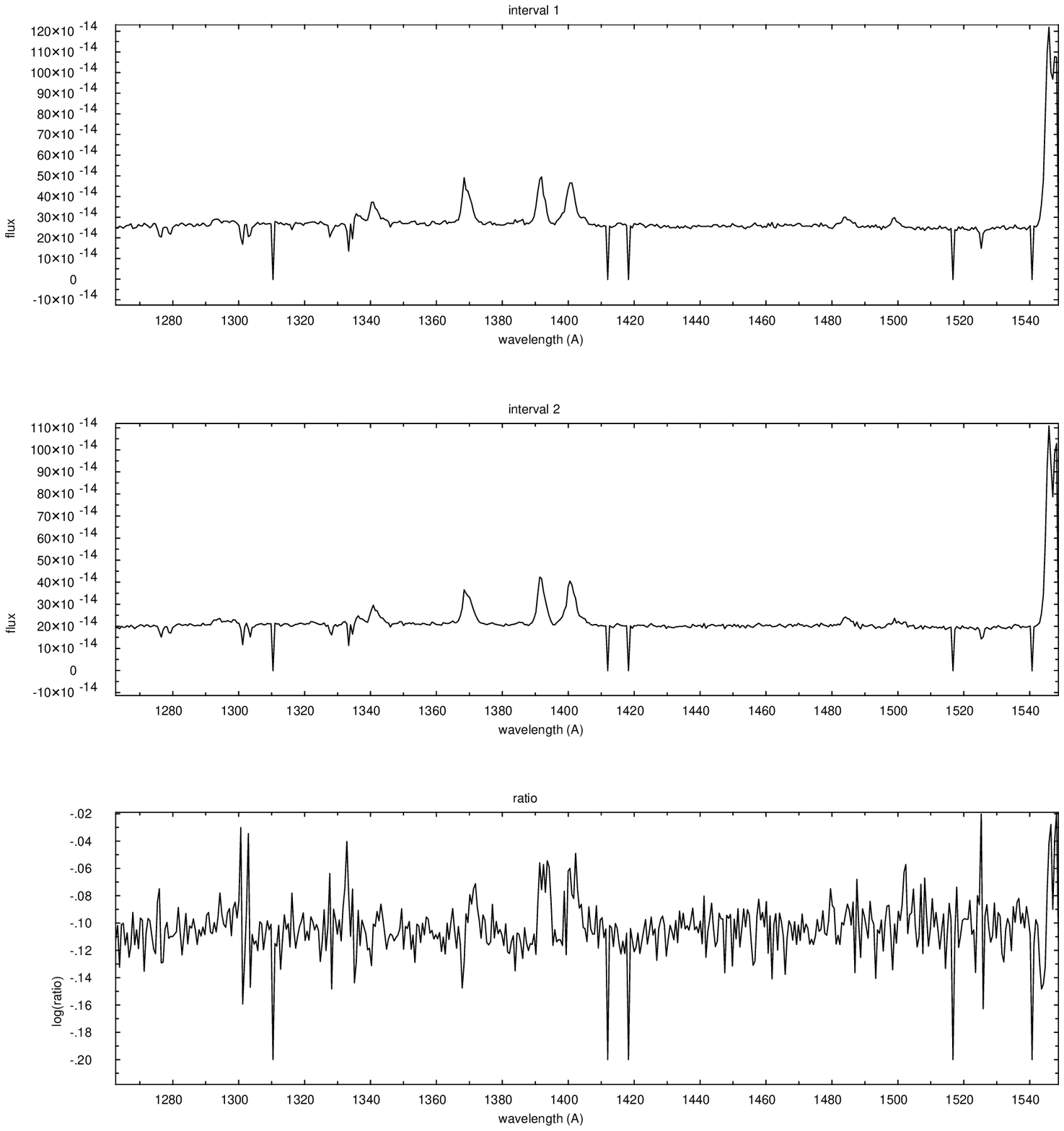} 
\epsfbox{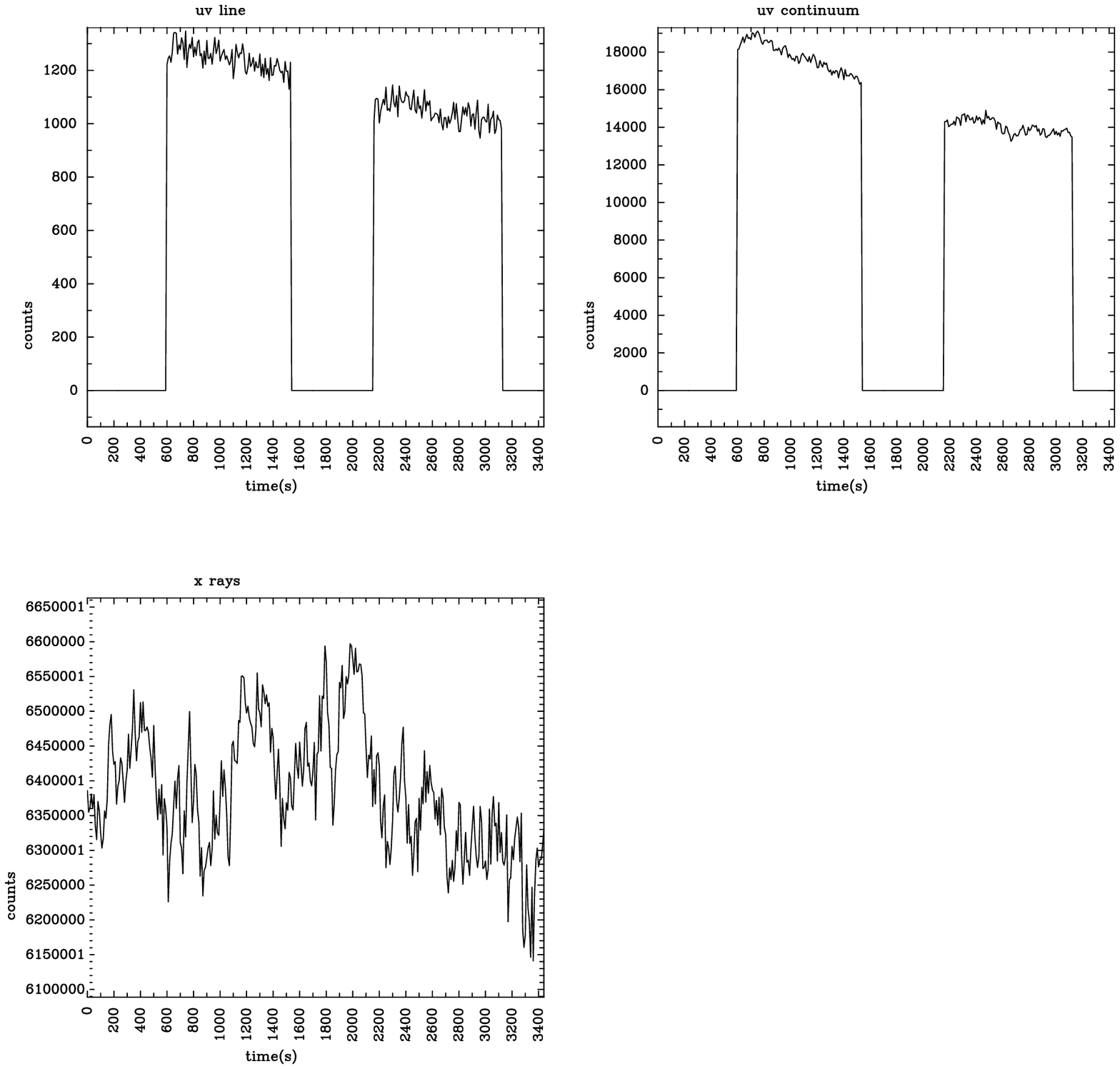} 
\epsfbox{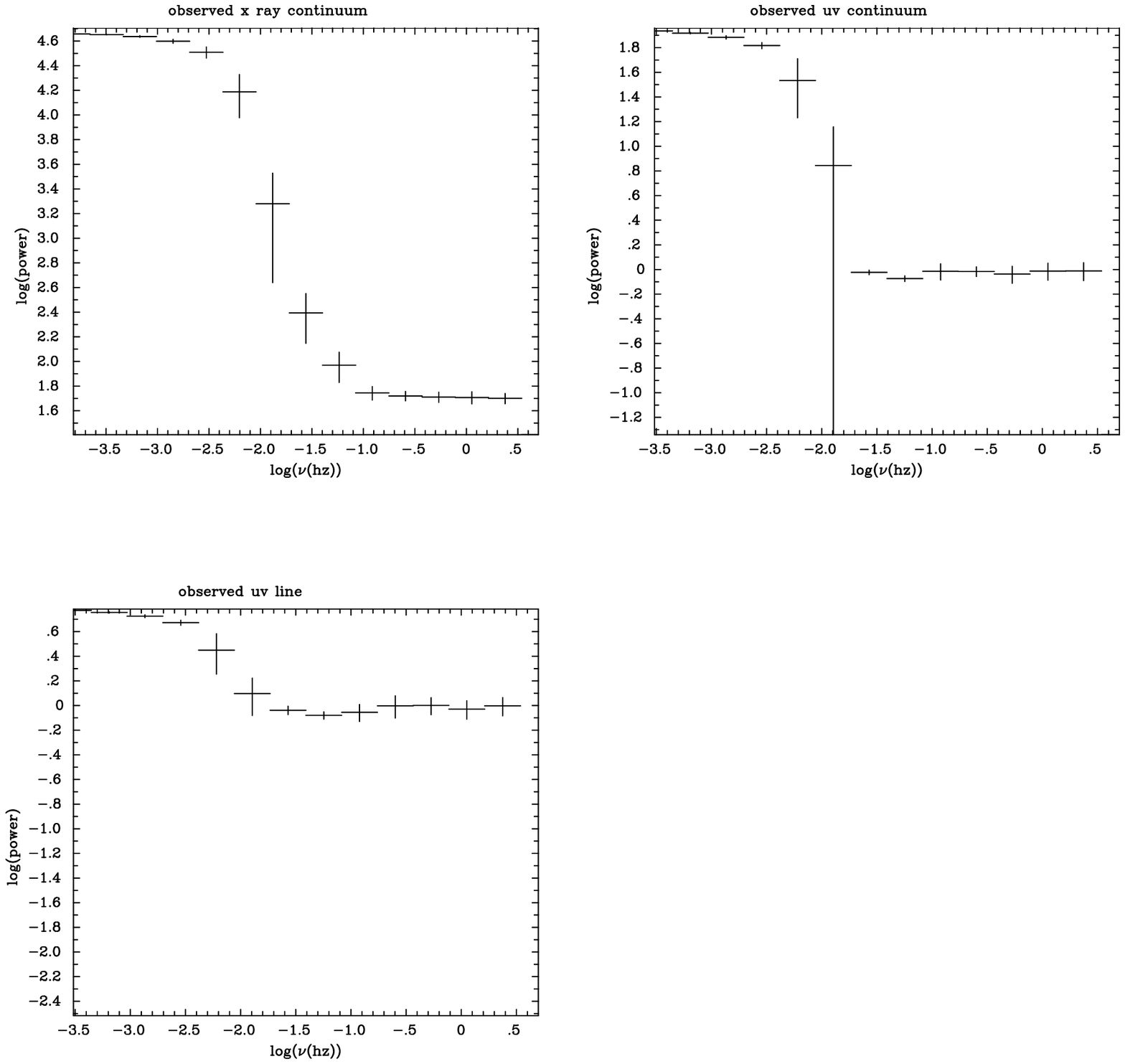} 
\epsfbox{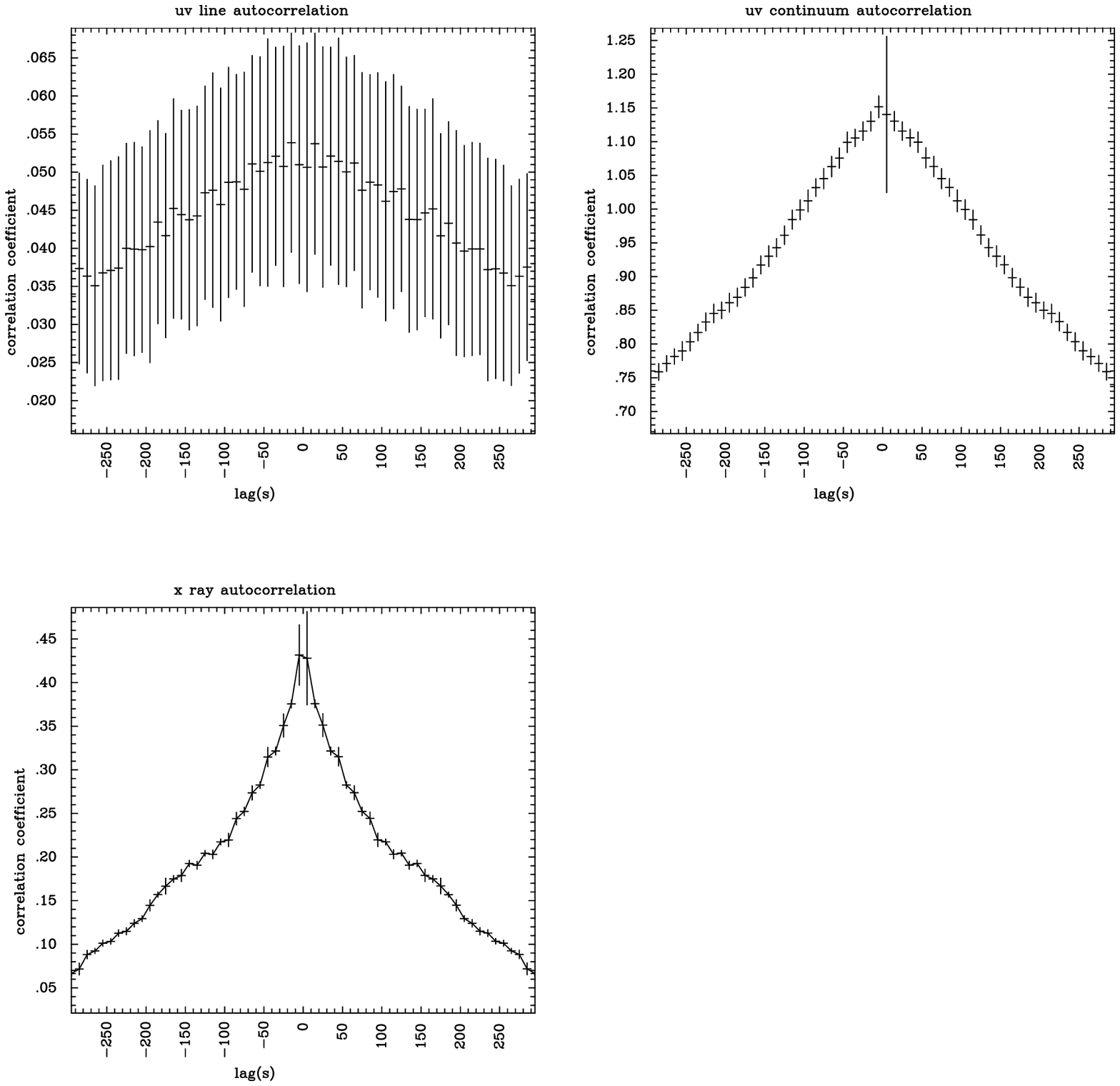} 
\epsfbox{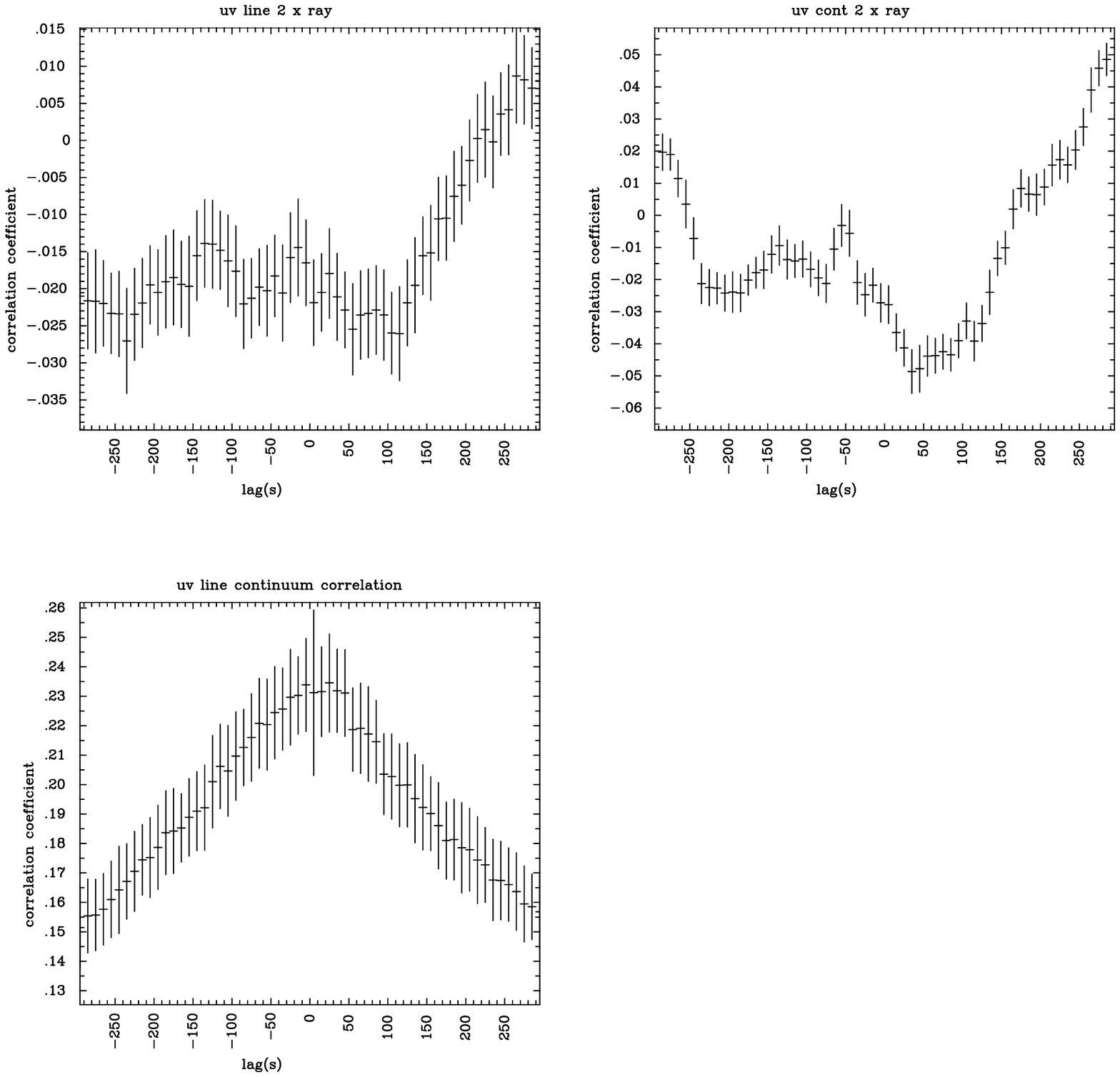} 
\epsfbox{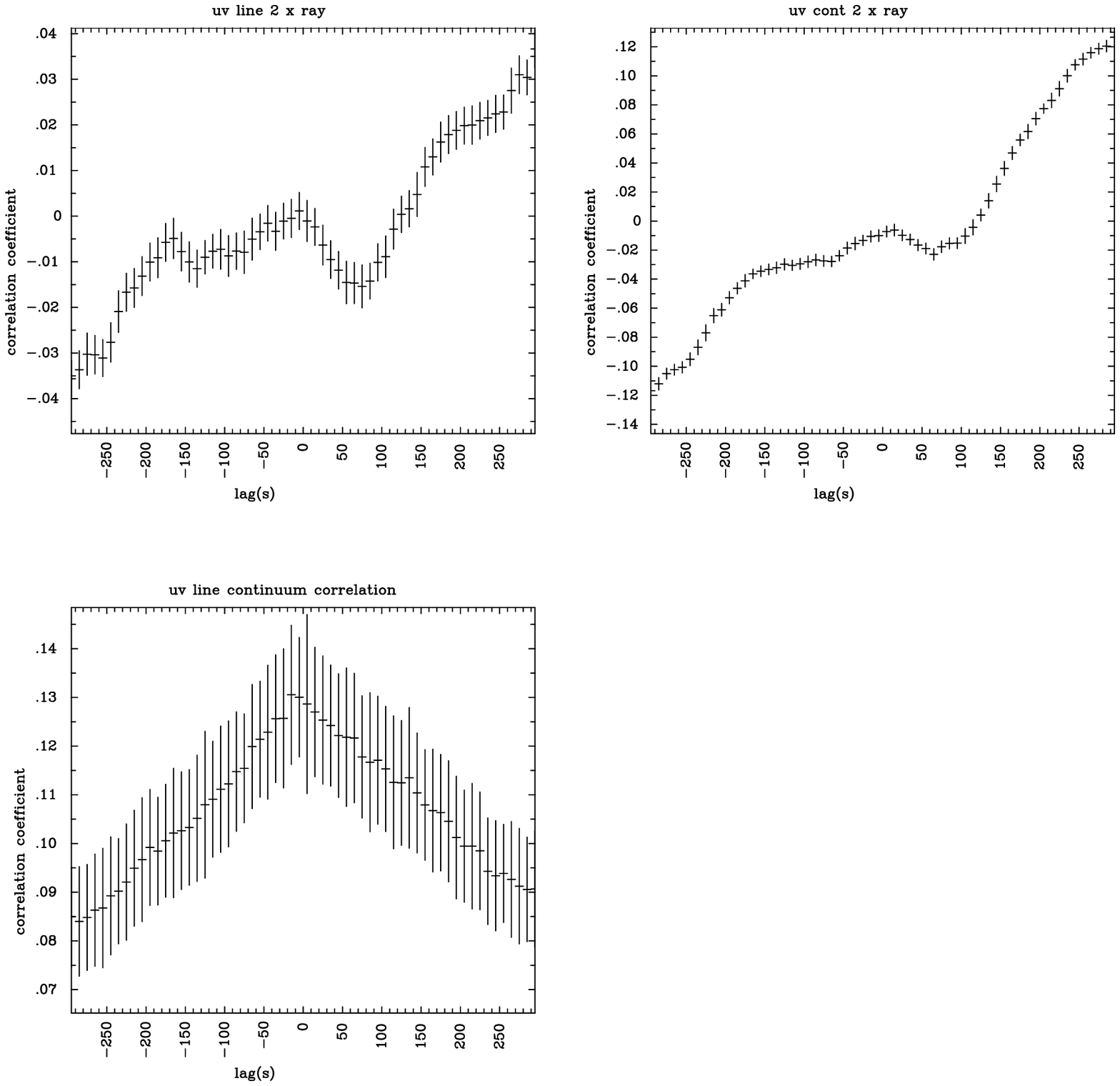} 
\epsfbox{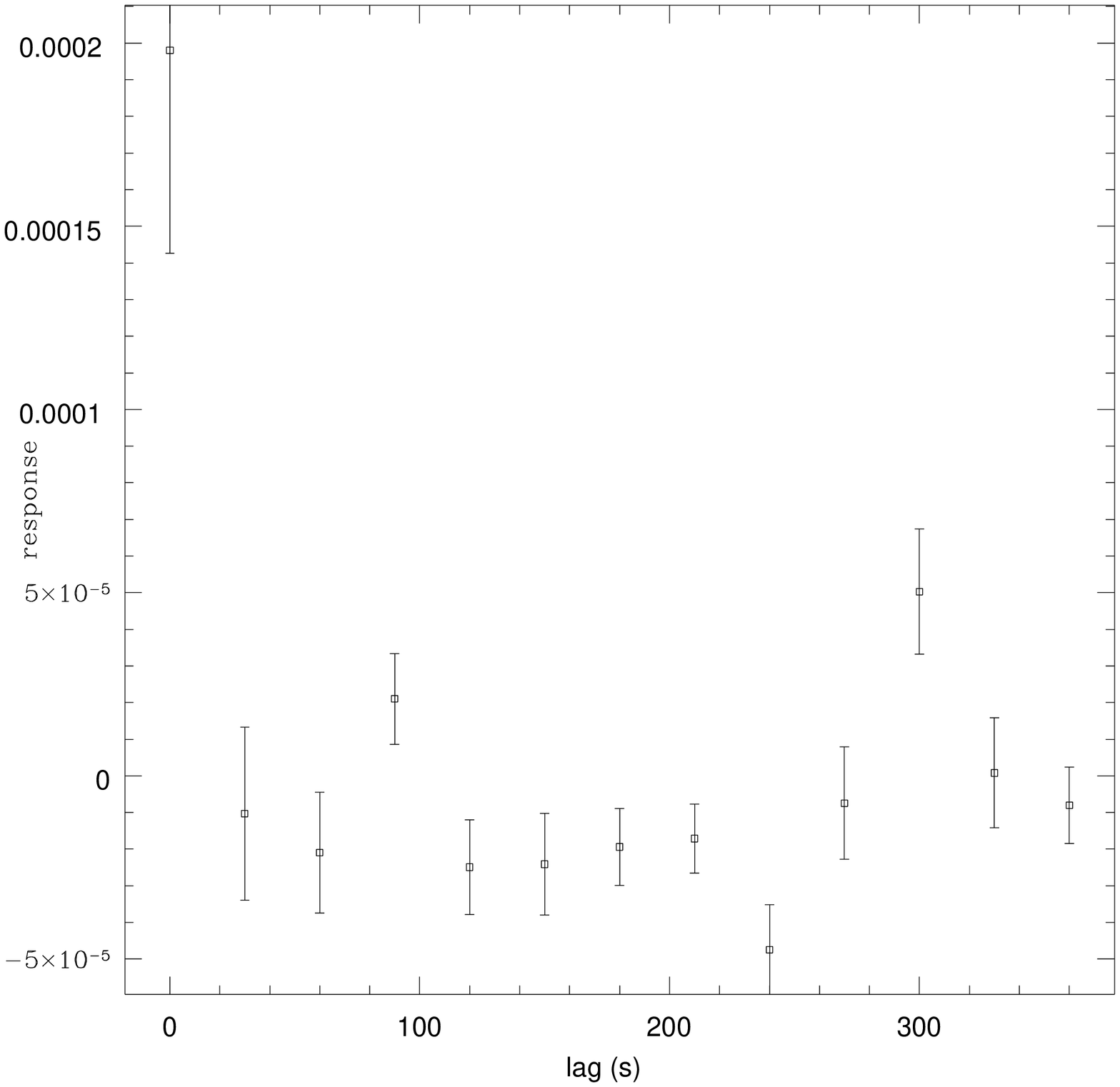} 
\epsfbox{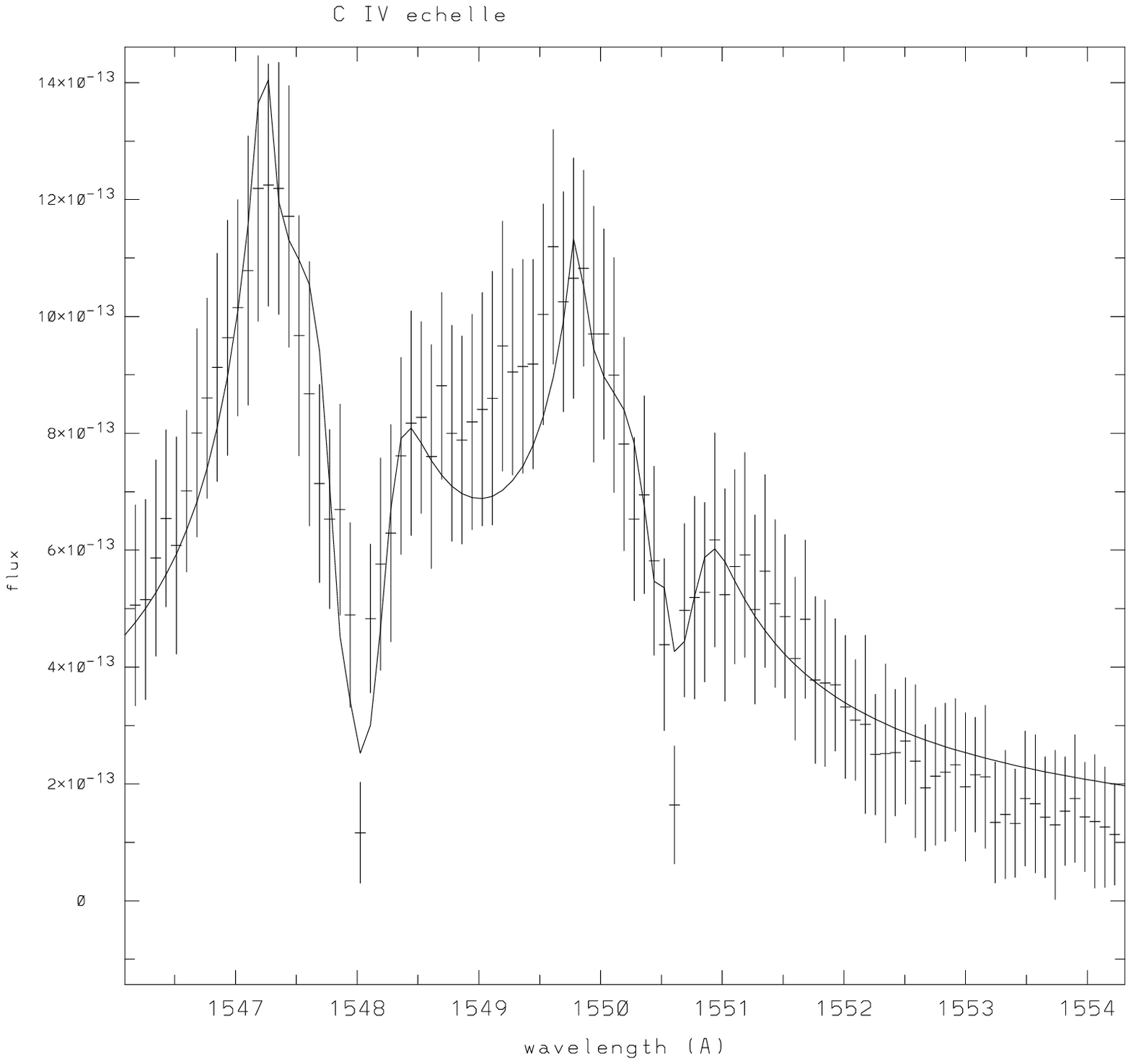} 
\epsfbox{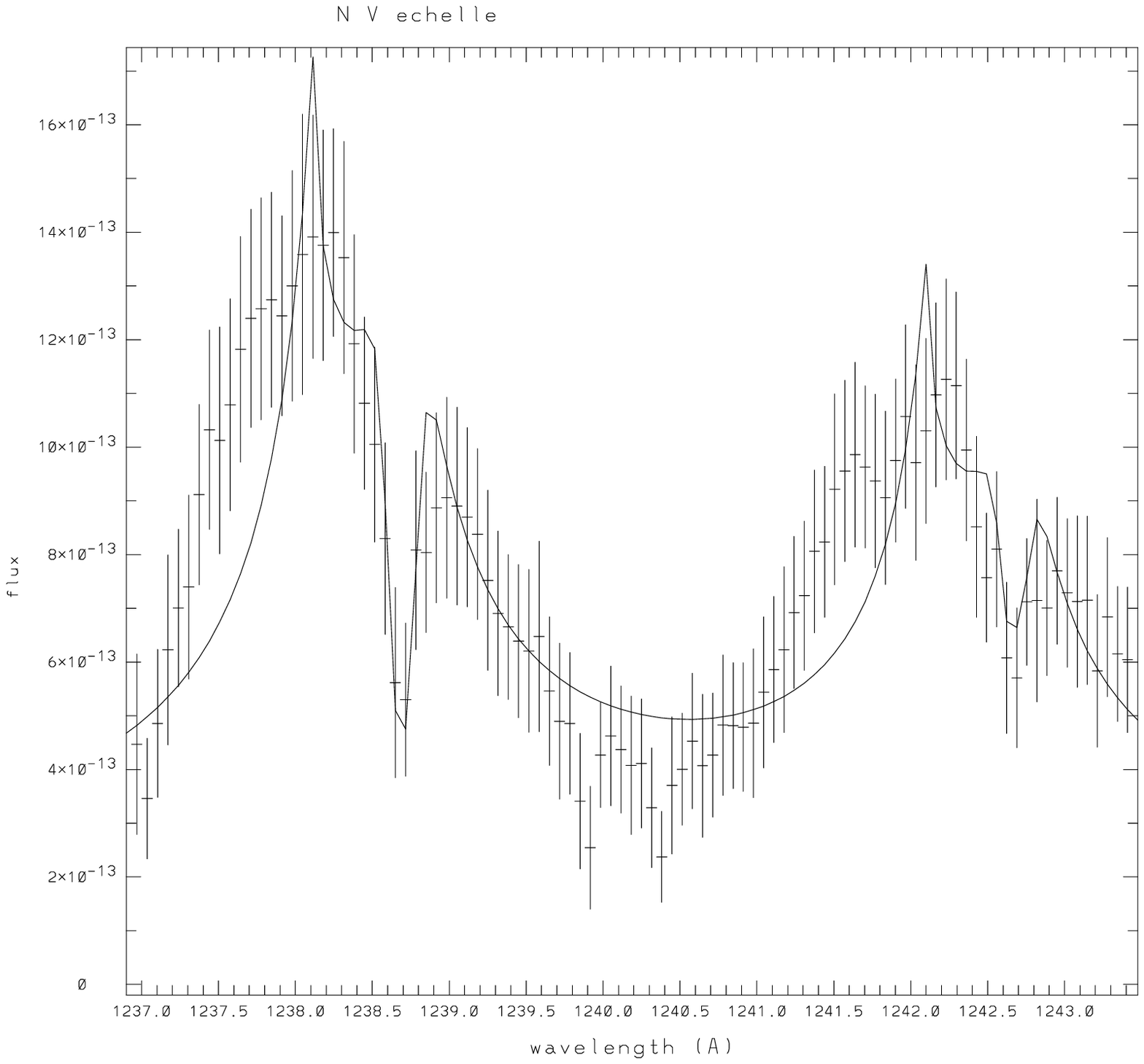} 
\epsfbox{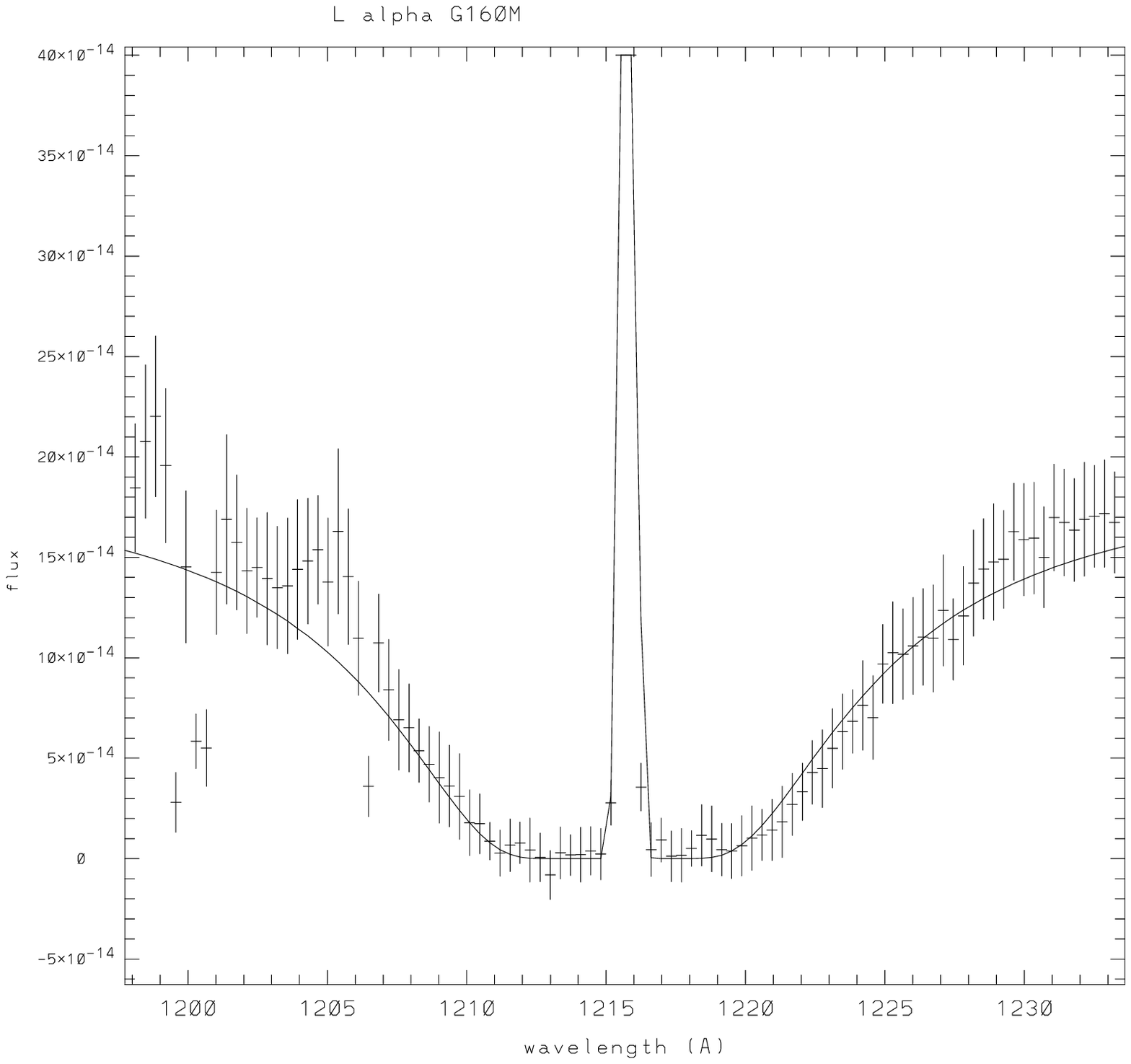} 

\end{document}